\documentclass[aps,pra,reprint,twocolumn,superscriptaddress,floatfix,nofootinbib]{revtex4-1}
\usepackage{epsfig,amsmath,amssymb,color,comment,physics, dsfont, bbold}
\usepackage[makeroom]{cancel}
\usepackage[caption=false]{subfig}
\usepackage{mathrsfs}
\usepackage[countmax]{subfloat}
\usepackage[normalem]{ulem}
\usepackage[english]{babel}
\usepackage{dsfont}
\usepackage{float}
\usepackage[bookmarks=true,colorlinks,linkcolor=blue,urlcolor=NavyBlue,citecolor=RoyalBlue]{hyperref}
\usepackage[dvipsnames]{xcolor}
\usepackage{color}

\newcommand{\sx}{\hat{\sigma}^x}
\newcommand{\sy}{\hat{\sigma}^y}
\newcommand{\sz}{\hat{\sigma}^z}
\newcommand{\Pz}{\hat{P}}

\graphicspath{{./Figures/}}

\begin{document}

\title{Deep thermalization in constrained quantum systems}

\author{Tanmay Bhore} 
\affiliation{School of Physics and Astronomy, University of Leeds, Leeds LS2 9JT, UK}

\author{Jean-Yves Desaules} 
\affiliation{School of Physics and Astronomy, University of Leeds, Leeds LS2 9JT, UK}

\author{Zlatko Papi\'c} 
\affiliation{School of Physics and Astronomy, University of Leeds, Leeds LS2 9JT, UK}

\date{\today}
\begin{abstract}
The concept of ``deep thermalization’’ has recently been introduced to characterize moments of an ensemble of pure states, resulting from projective measurements on a subsystem, which lie beyond the purview of conventional Eigenstate Thermalization Hypothesis (ETH). In this work, we study deep thermalization in systems with kinetic constraints, such as the quantum East and the PXP models, which have been known to weakly break ETH by slow dynamics and high sensitivity to the initial conditions. We demonstrate a sharp contrast in deep thermalization between the first and higher moments in these models by studying quench dynamics from initial product states in the computational basis: while the first moment shows good agreement with ETH,  higher moments deviate from the uniform Haar ensemble at infinite temperature. We show that such behavior is caused by an interplay of time-reversal symmetry and an operator that anticommutes with the Hamiltonian. We formulate sufficient conditions for violating deep thermalization, even for systems that are otherwise ``thermal'' in the ETH sense. By appropriately breaking these properties, we illustrate how the PXP model fully deep-thermalizes for all initial product states in the thermodynamic limit. Our results highlight the sensitivity of deep thermalization as a probe of physics beyond ETH in kinetically-constrained systems.
\end{abstract}
\maketitle

\section{Introduction}

Thermal equilibrium is the eventual fate of an isolated chaotic quantum system in the absence of special mechanisms such as integrability~\cite{Sutherland} or localization~\cite{Huse-rev,AbaninRev}. This behavior is encapsulated by the Eigenstate Thermalization Hypothesis (ETH) \cite{Srednicki96, DeutschETH}, which explains thermalization at the level of subsystems: under unitary dynamics, a subsystem entangles with its complement such that its reduced density matrix increasingly resembles a thermal Gibbs state. The complementary subsystem, whose states are traced over, assumes the role of thermal bath during the unitary dynamics. This scenario has been extensively tested numerically~\cite{RigolNature} and in experimental setups of quantum simulators with cold atoms, trapped ions and other engineered quantum systems~\cite{ETH_review_Alessio_2016, Gogolin2016, Ueda2020}. 

\begin{figure}[t!]
	\centering
	\includegraphics[width=\linewidth]{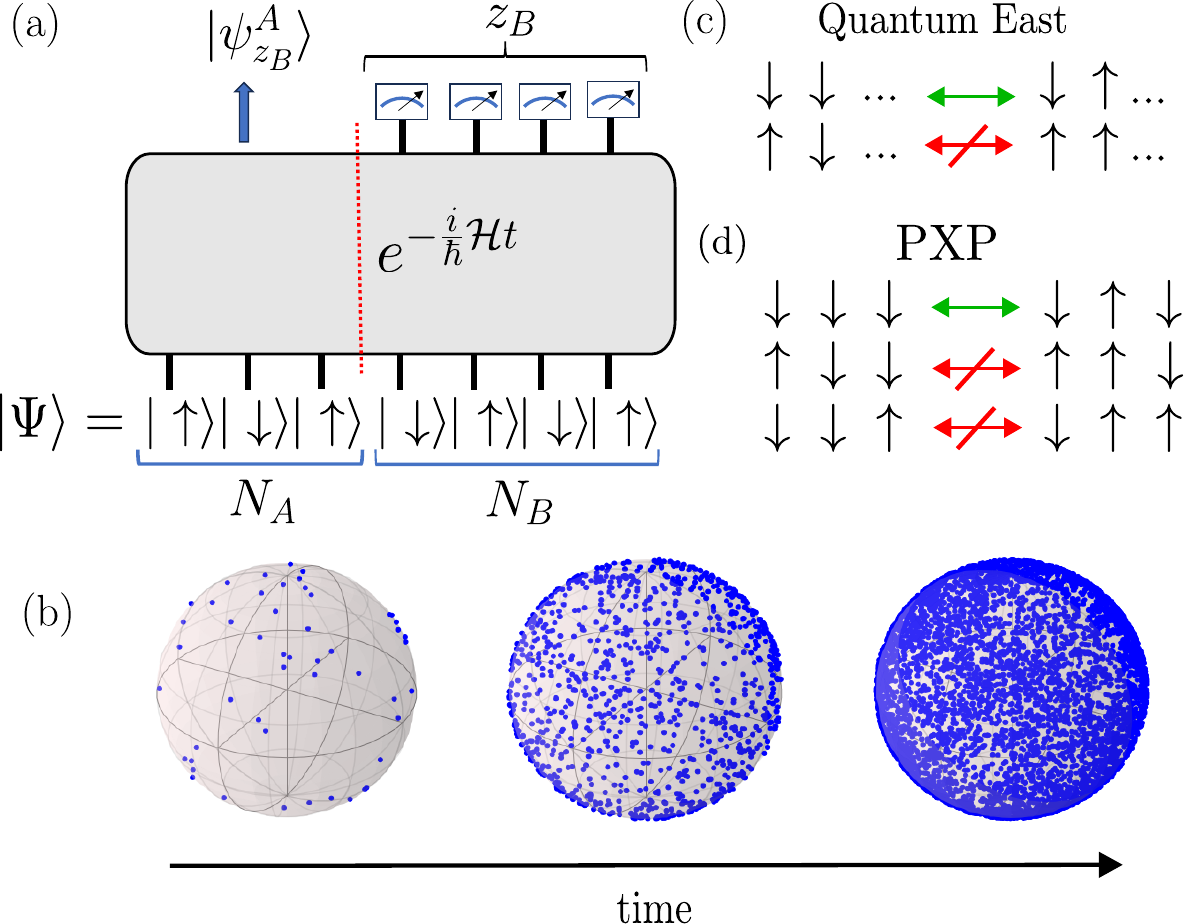}
	\caption{ (a)-(b): The projected ensemble. A typical setup consists of a spin system, prepared in a product state. The system then evolves under unitary dynamics generated by some Hamiltonian $\mathcal{H}$, which keeps it in a pure state $\ket{\psi}$. After time $t$,  projective measurements are performed on each spin in the subsystem $B$. The output of this measurement, $\ket{z_B}$, indexes the resulting pure state on subsystem $A$, labeled by $\ket{\psi_{z_B}^{A}}.$  For all the models studied in this work, we pick the subsystem $A$ as the first $N_A$ spins, and subsystem $B$ as the remaining $N{-}N_A$ spins.
    (b) The evolution of states in the projected ensemble for a single qubit. With time, the states uniformly occupy the Bloch sphere, becoming indistinguishable from the Haar ensemble.
    (c)-(d): The constrained models considered in this work are defined in terms of Pauli $x$-operators flipping a spin, subject to the state of its neighbor(s). In the quantum East model (c), the spin flip is allowed (green arrow) if the left neighbor is in the $\downarrow$ state, regardless of the state of right neighbor. In the PXP model (d), both neighbors must be in the $\downarrow$ state for a flip to occur. From these rules, it is clear that if the spin in $B$, adjacent to the bipartition, is in $\uparrow$ state, the state of $A$ cannot be fully random. This must be taken into account with an appropriate postselection rule, as discussed in Sec.~\ref{sec:syms}.
    }
	\label{fig:illustration_PE}
\end{figure}

The conventional picture of ETH, however, is blind to the microscopic details of the bath. Recent progress in the control and manipulation of individual degrees of freedom in quantum simulators~\cite{Bloch2012, Kjaergaard, MonroeRMP, Browaeys2020} has brought in a refinement of this picture: states of the bath can be explicitly measured and the state of the subsystem can be studied \textit{conditional} to a measurement outcome on the bath~\cite{Choi_nature_2023,Cotler2023}. Repeating such projective measurements in a fixed, local basis gives rise to an ensemble of pure states on the subsystem which, along with their corresponding Born probabilities, form the so-called ``projected ensemble''~\cite{Popp2005,goldstein2016}. 
Through quench dynamics experiments on Rydberg atom arrays~\cite{Choi_nature_2023}, numerical simulations of Hamiltonian models~\cite{Cotler2023, Lucas2023} and Floquet circuits~\cite{Wen_Choi_2022_PRL,Ippoliti_Wen_QSD_2023}, universal behavior has been found in the dynamics of the projected ensemble: under chaotic time evolution at infinite temperature, the statistical properties of the projected ensemble become indistinguishable from a uniform ensemble on the Hilbert space, i.e., the maximally entropic Haar ensemble~\cite{Harrow_Haar_2013}. Remarkably, \emph{all} moments of the projected ensemble were argued to become indistinguishable from those of the Haar ensemble. In the language of quantum information theory, the moments then furnish (approximate) quantum state designs~\cite{renes2004symmetric,ambainis2007}. The picture summarized above, dubbed ``deep thermalization''~\cite{Choi_nature_2023,Cotler2023},  illuminates a distinct role of measurements compared to a simple hindrance to thermalization. Accordingly, ETH is generalized to the statistical distribution of wave functions, instead of just expectation values of physical observables. This is part of a broader push to probe physics beyond ETH, complementing other approaches such as free probability~\cite{Silvia_ETH_PRL_2023, Silvia_free_cumulants_2023}, which make predictions about higher-order correlation functions.

The practical appeal of deep thermalization in chaotic systems lies in its universality: Haar-random ensembles can be generated under unitary dynamics starting from a simple initial state~\cite{Choi_nature_2023}. An intriguing question therefore arises how this phenomenology changes in chaotic systems whose thermalization dynamics is strongly dependent on the initial state, such as systems displaying Hilbert space fragmentation, quantum many-body scars and other types of ``weak'' ergodicity breaking~\cite{Serbyn2021, MoudgalyaReview, ChandranReview}. The paradigmatic examples include the experimental systems of Rydberg atoms~\cite{Bernien2017, Bluvstein2021} and ultracold bosons in a tilted optical lattice~\cite{GuoXian2022}, which feature persistent quantum revivals when quenched from a special initial state, while they quickly thermalize for typical initial conditions. Such systems are described by an effective spin model called the ``PXP model"~\cite{FendleySachdev, Lesanovsky2012}, which imposes a kinetic constraint on simultaneous flips of neighboring spins. Similar types of constraints can give rise to slow glassy dynamics in the quantum East model~\cite{vanHorssen2015,Pancotti2020}. In this paper we explore the nature of deep thermalization and its sensitivity to initial conditions for such constrained systems.  

In both the PXP and quantum East models studied below, we find deep thermalization to be absent, even at infinite temperature, while the first moment of the projected ensemble agrees well with ETH. We demonstrate this for a large class of time-evolved initial states as well as the eigenstates of the models. We elucidate the origin of this surprising  behavior, finding that it does not stem from the constraints but rather from spectral properties and the existence of special operators that anticommute with the Hamiltonian.  Once these are properly taken care of, we observe a restoration of deep thermalization in the thermodynamic limit. This is true even in the PXP model, where clear signs of ergodicity breaking are found in all accessible finite sizes.
Our results illustrate that deep thermalization is not only a hallmark of ``maximally chaotic'' models but more broadly present in models that also display weak ergodicity breaking.  The results furthermore highlight the sensitivity of deep thermalization framework compared to standard ETH, as the former requires more care beyond resolving the usual global symmetries of the model. 

This paper is organized as follows: In Sec. \ref{sec:projectedensemble}, we review the construction of the projected ensemble and how we quantify its distance from the Haar ensemble. We also discuss the crucial role of symmetries, which are illustrated for the Sachdev-Ye-Kitaev (SYK) model in Sec.~\ref{sec:SYK}. This maximally-chaotic model will be used as a benchmark for the rest of the paper. In Sec.~\ref{sec:t_sym}, we show how time-reversal symmetry can prevent deep thermalization of eigenstates, and we illustrate this using the Ising model in mixed transverse and longitudinal fields. 
As our first kinetically constrained model, we consider the quantum East model in its thermalizing phase in Sec.~\ref{sec:PX}.  In Sec.~\ref{sec:antisym}, we show how the interplay of time-reversal invariance and ``antisymmetries'' can generally prevent deep thermalization for time-evolved states. Finally, in Sec.~\ref{sec:PXP}, we study deep thermalization in the constrained PXP model describing one-dimensional Rydberg atom arrays. Here we focus in particular on different types of initial states that have been known to give rise to anomalous dynamical behavior associated with quantum many-body scars~\cite{Bernien2017, Turner2018b}. Our Conclusions are presented in Sec.~\ref{sec:conclusion}, while Appendixes contain further details of the analysis and effect of perturbations to the PXP model.

\section{The projected ensemble}\label{sec:projectedensemble}

Consider a quantum spin system in a pure state $\ket{\Psi}$. The state can be prepared, e.g., via unitary time evolution from some product state, as sketched in Fig.~\ref{fig:illustration_PE}(a). The system is bipartitioned into two contiguous regions: a subsystem $A$ and its complement $B$, consisting of $N_A$ and $N_B$ spins, respectively. For simplicity, we assume the total Hilbert space is given by the tensor product $\mathcal{H}_A \otimes \mathcal{H}_B$, although, as explained in Sec.~\ref{sec:syms}, this assumption can be relaxed in certain cases. We consider performing projective measurements on the subsystem $B$ in the local computational basis, ${\ket{z}}$. Each such measurement outputs a classical string $z_B$ of length $N_B$. According to the Born rule, such a measurement leaves the subsystem $A$ in a pure state which is conditional to measuring the string $z_B$ in the complementary subsystem $B$, with a probability $p_{z_B}$ given by: 
\begin{align}
p_{z_B} &= \bra{\Psi} (\mathbb{1}_A \otimes \ket{z_B} \bra{z_B} ) \ket{\Psi}, \\
\ket{\psi^A_{z_B}} &=   \frac{1}{\sqrt{ p_{z_B}} } (\mathbb{1}_A \otimes \bra{z_B} ) \ket{\Psi}.
\end{align}
Note that each pure state on $A$ is indexed by a measurement outcome  $z_B$ on $B$. With the states $\ket{\psi^A_{z_B}}$ and probabilities $p_{z_B}$, the projected ensemble is defined as the set:
\begin{equation}
    \mathcal{E} = \{ p_{z_B}, \ket{\psi^A_{z_B}} \}.
\end{equation}
Since the measurement basis $\{ \ket{z_B}\}$ is orthonormal, the probabilities $\{ p_{z_B} \}$ sum to unity. The projected ensemble then forms a probability distribution on $\mathcal{H}_A$, whose $k^{th}$ moment is given by:
\begin{equation}\label{eq:probdistribution}
    \rho_{\mathcal{E}}^{(k)} = \mathbb{E}_{\psi \sim \mathcal{E}} \left[ ( \ket{\psi}\bra{\psi} )^{\otimes k}  \right] = \sum_i p_i (\ket{\psi_i}\bra{\psi_i})^{\otimes k},
\end{equation}
where the sum runs over all states in the projected ensemble $\mathcal{E}$ and $\rho_{\mathcal{E}}^{(k)}$ acts as a density operator on the Hilbert space $\mathcal{H}_A^{\otimes k}$, corresponding to $k$ replicas of $\mathcal{H}_A$.

The first moment of the distribution in Eq.~(\ref{eq:probdistribution}) is exactly the reduced density matrix on $\mathcal{H}_A$:
\begin{equation}
 \rho_{\mathcal{E}}^{(1)} = \sum_i p_i \ket{\psi_i}\bra{\psi_i} 
 = {\rm Tr}_B(\ket{\Psi}\bra{\Psi}),   
\end{equation}
which is the central object of study in ETH. However, higher moments of the projected ensemble $(k{>}1)$ are in general not equal to $k-$fold tensor products of the reduced density matrix.  
As such, the projected ensemble strictly encodes more information than the reduced density matrix on $\mathcal{H}_A$. 

\subsection{Haar ensemble}\label{sec:haar}

Consider the time evolution of an isolated quantum system initialized in a far-from-equilibrium state. ETH tells us that for ergodic systems in the absence of conservation laws, the reduced density matrix $\rho_A(t)$ of a subsystem $A$ (much smaller than the rest of the system) relaxes to a thermal Gibbs state at late times, i.e., at times much larger than the inverse of the system's microscopic energy scales. This can be formally expressed as $\lim_{t \rightarrow \infty} \rho_A(t) = \exp(-\beta \hat{H})/\mathcal{Z}$, where $\beta$ is the inverse temperature, set by the energy of the initial state, and $\mathcal{Z}$ is a normalizing factor. At finite times and finite system sizes, the equality is only exact up to fluctuations that typically subside exponentially with system size. During the dynamics, the microscopic details of the initial state are progressively  lost and the system increasingly resembles a featureless thermal state.

Since the first moment of the projected ensemble is exactly the reduced density matrix, it is natural to ask whether this evolution to a featureless thermal state holds for all moments of the projected ensemble. We note that the evolution to a thermal Gibbs state is guided by the principle of maximization of entropy, i.e., the second law of thermodynamics. However, since we are dealing with a probability distribution, it is natural to consider maximizing the entropy of this distribution. This is accomplished by conjecturing that the projected ensemble dynamically evolves to the maximally entropic uniform distribution of pure states on a Hilbert space. For initial states with energies in the middle of the spectrum, \textit{i.e.} states at infinite temperature, this maximally entropic ensemble is the Haar ensemble~\cite{Harrow_Haar_2013}. One can then quantify the ``depth'' of thermalization by testing to what extent the moments of the projected ensemble agree with those of the Haar ensemble, see Fig.~\ref{fig:illustration_PE}(b). 

For pure states $\ket{\psi}$ in a Hilbert space, $k^{th}$-moments of the Haar ensemble can be constructed as:
\begin{equation}
    \rho_\mathrm{Haar}^{(k)} = \int {\rm d} \psi ( \ket{\psi}\bra{\psi} )^{\otimes k},
\end{equation}
where the integral runs over the entire Hilbert space $\mathcal{H}$. The $k$-th moment admits an analytical expression~\cite{Harrow_Haar_2013,Cotler2023}
\begin{equation}
    \rho_\mathrm{Haar}^{(k)} = \frac{\Pi_k}{\binom{d+k-1}{k}},
\end{equation}
where $\Pi_k$ is the subspace of $\mathcal{H}^{\otimes k}$ invariant under all permutations of the $k$ copies and $d$ is the dimension of $\mathcal{H}$.
\\ As a consistency check of the conjecture, we highlight that the first moment (mean) of the Haar ensemble is the Identity operator $\hat{\mathds{I}}/d$, which is exactly equal to the Gibbs ensemble at infinite temperature. At finite temperatures, the existence of a universal random ensemble has been hinted at, but its exact form remains unknown. \cite{Cotler2023}. Here on, we restrict our focus to infinite temperature initial states.
\\ To quantify the degree to which a system deep thermalizes, the trace distance between the $k^{th}$-moments has been used in the literature~\cite{Cotler2023,Choi_nature_2023}: 
\begin{equation}
    \Delta^{(k)} = \frac{1}{2} \norm{ \rho_{\mathcal{E}}^{(k)} - \rho_\mathrm{Haar}^{(k)} }_1, 
\end{equation}
where $\norm{.}_1$ denotes the trace norm. Note that $\Delta^{(k)}$ follows a monotonicity relation such that $\Delta^{(k')} < \Delta^{(k)}~\forall k'<k$. This is closely linked to the concept of quantum k-designs, an important resource in quantum information theory. We say an ensemble $\mathcal{E}$ forms a $k$-design if $\Delta^{(k)}=0$, which implies that $\Delta^{(k')} = 0, ~\forall k'<k$ . Then, a system deep thermalizes for a given $k$ if it forms a $k$-design in the thermodynamic limit. The way this limit is taken is by fixing the subsystem $A$ and then increasing the size of the ``reservoir'' subsystem $B$. Since we are limited to finite-sized systems in this study, we study the scaling of $\Delta^{(k)}$ as $N_B$ is increased. As shown in Ref.~\cite{Cotler2023}, for a chaotic system $\Delta^{(k)}$ is expected to decrease exponentially in $N_B$ for all $k$. The rate of the exponential decay, however, can be different for different $k$ values and, typically, larger $k$ are more strongly impacted by finite size effects.

\subsection{Effect of symmetries and constraints}\label{sec:syms}

Until this point, we have considered the simple case where energy is the only conserved quantity. When additional conservation laws are present, generic states instead thermalize to a Generalized Gibbs Ensemble~\cite{Essler2016}. This ensemble now features Lagrange multipliers enforcing the conservation of all charges. As a consequence, the projected ensemble no longer approaches the Haar ensemble,  but a similar generalized Haar ensemble can be introduced using the same principle~\cite{Lucas2023}.
This construction depends on the specifics of the model and its conserved charges, meaning that it generally has a much more complicated structure than the Haar ensemble. In order to simplify the study of deep thermalization with conservation laws, one can restrict to a single charge sector. However, this is usually not sufficient to recover thermalization to the Haar ensemble, as the conserved charge can introduce correlations between the measured string $z_B$ and the projected state $\ket{\psi^A_{z_B}}$. To recover the Haar ensemble, one then needs to use \textit{postselection} on the measured strings $z_B$~\cite{Cotler2023}. We present an example of this procedure in the following section.

The need for postselection also arises in a conceptually different case when dynamical constraints are present. The models we focus on in this work (Sec.~\ref{sec:PX} and \ref{sec:PXP}) are defined in terms of Pauli $x$-matrices, which allow each spin to flip its state. However, the spin-flip operators are dressed with projectors, which encode dynamical constraints: a spin can only flip its state if its immediate neighbors to the left or right are in the $\downarrow$ state, see Figs.~\ref{fig:illustration_PE}(c)-(d). While such a constraint cannot be expressed as a local operator that commutes with the Hamiltonian, it similarly places strong restrictions on the dynamics. For example, if the boundary spin in $B$ subsystem (i.e., the one closest to the $A$ subsystem) is in the $\uparrow$ state, this will impact the state of the spin in $A$ closest to the bipartition. This will be true at all evolution times, hence the state of $A$ will never reach the Haar ensemble. Nevertheless, since the constraints considered in this work are all local and affect only the nearest-neighbor spin pairs, it is easy to take care of them via postselection by keeping only the measurement outcomes in which the boundary spin in $B$ is in the $\downarrow$ state. This will, \emph{a priori}, not exclude any configurations in $A$, hence deep thermalization becomes possible, as we show in Secs.~\ref{sec:PX}-\ref{sec:PXP} below. For a constrained system such as the PXP model in Fig.~\ref{fig:illustration_PE}(d), the total Hilbert space does not have a tensor product structure, which can be seen from the fact that the total Hilbert space dimension grows as a Fibonacci number~\cite{Feiguin07}. Thus, with the help of postselection, we can lift the assumption that the total Hilbert space must be written in the form $\mathcal{H}_A \otimes \mathcal{H}_B$.

\section{Maximally chaotic SYK model}\label{sec:SYK}

A natural toy model for illustrating the concept of deep thermalization is the Sachdev-Ye-Kitaev model~\cite{SachdevYe93,Kitaev15,rosenhaus2019}. This model serves as a paradigm of quantum many-body chaos \cite{Kobrin_SYK_PRL_2021}, thus we expect it to yield good agreement between the projected and Haar ensembles. The model is formulated in terms of $2N$ Majorana fermions with all-to-all interactions, 
\begin{equation}\label{eq:SYK_mf}
    \hat{H}_\mathrm{SYK}=\sum_{i<j<k<l} J_{ijkl} \hat{\chi}_i\hat{\chi}_j\hat{\chi}_k\hat{\chi}_l,
\end{equation}
where $\hat{\chi}$ denote the Majorana operators and the summation indices take values between 1 and $2N$. The couplings $J_{ijkl}$ are randomly drawn from a normal distribution with mean 0 and variance $6/(2N)^3$. One of the many interesting features of the SYK model is that its random-matrix theory (RMT) class changes with system size~\cite{You2017}. Indeed, for $N{=}4n$ the model exhibits spectral statistics of the Gaussian Orthogonal Ensemble (GOE), $N{=}4n+1$ and $N{=}4n+3$ correspond instead to the Gaussian Unitary Ensemble (GUE), and for $N{=}4n+2$ to the Gaussian Symplectic Ensemble (GSE)~\cite{Mehta2004}. 

\begin{figure}[tb]
	\centering
	\includegraphics[width=\linewidth]{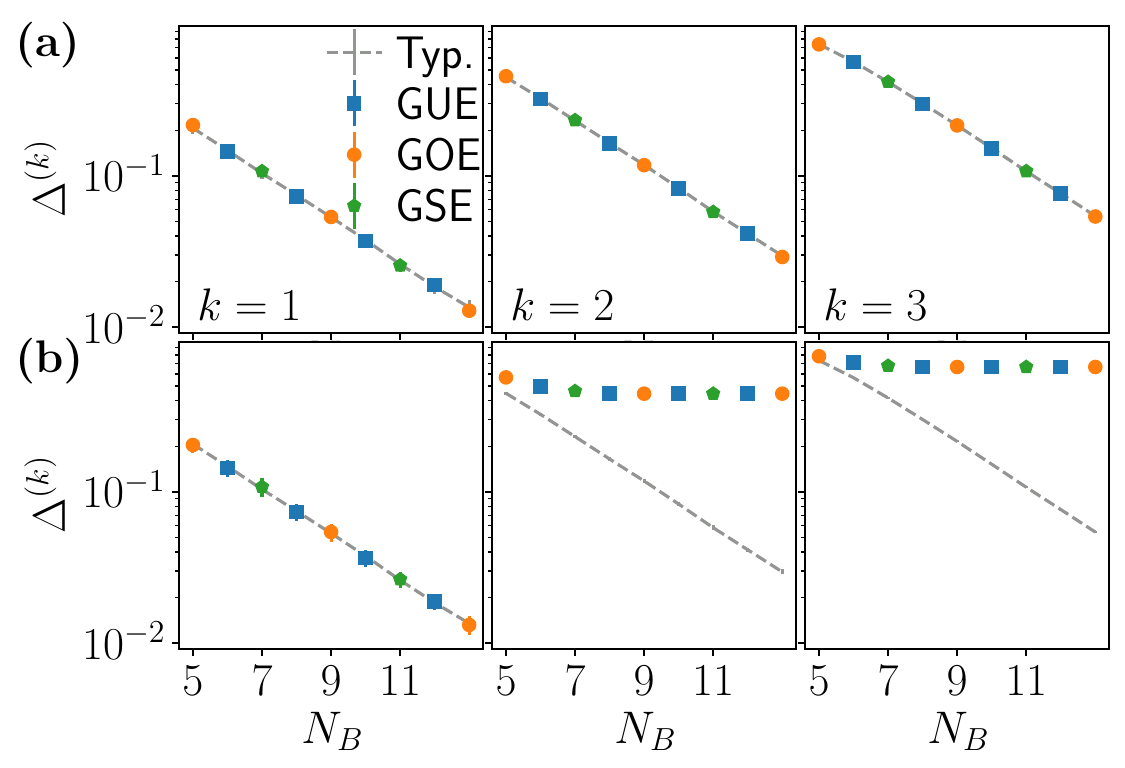}
	\caption{Late-time average of $\Delta^{(k)}$ in the SYK model with $N_A{=}4$ and various system sizes when starting from (a) state in Eq.~(\ref{eq:bloch}) with random angles, and (b) a computational basis state. For each system size, the data is for a \emph{single} random realization as there is no visible variation between them.
    }
	\label{fig:SYK_full}
\end{figure}

For the numerical study, it is more convenient to work with spin-1/2 operators. We use the Jordan-Wigner transformation, 
\begin{equation}
    \sqrt{2}\hat{\chi}_i =\begin{cases}
    \left(\prod_{j<k} \hat{\sigma}^z_j \right) \hat{\sigma}^x_k \quad i \ \text{even} \\
    \left(\prod_{j<k} \hat{\sigma}^z_j \right) \hat{\sigma}^y_k \quad i \ \text{odd} 
    \end{cases}
\end{equation}
to convert Eq.~(\ref{eq:SYK_mf}) to a model defined in terms of $N$ spin-1/2 degrees of freedom. The resulting Hamiltonian has a rather complicated structure due to the Jordan-Wigner strings, hence we do not write it down explicitly. Nonetheless, we note that the various terms in the model can be either purely diagonal, flip two spins, or flip four spins. This depends on how many of the Majorana fermions correspond to the same spin site. As a consequence, the spin model conserves the parity of the number of $\downarrow$ spins:
\begin{equation}
    \hat{Z}_P=\prod_{j=1}^N \sz_j.
\end{equation}

To probe deep thermalization in the SYK model, similar to Refs.~\cite{Cotler2023, Choi_nature_2023}, we study a global quench by preparing the system in different kinds of initial states. One natural choice is to use initial states belonging to the computational basis, which is also the measurement basis. As a second choice, we use product states where each spin $j$ points in a random direction on the Bloch sphere, 
\begin{equation}\label{eq:bloch}
\ket{\theta, \phi} \equiv \bigotimes_j \left( \cos\left(\theta_j/2\right)\ket{\uparrow_j}+e^{i\phi_j}\sin  \left(\theta_j/2\right)\ket{\downarrow_j}\right),
\end{equation}
where $\theta_j$ and $\phi_j$ are randomly drawn from the intervals $[0,\pi]$ and $[0,2\pi]$, respectively. Below we show that these two classes of initial states lead to stark differences in the deep thermalization.
Once we prepare the initial states, we evolve them using the SYK Hamiltonian and compute $\Delta^{(k)}$ at various time steps.

\begin{figure}[tb]
	\centering
	\includegraphics[width=\linewidth]{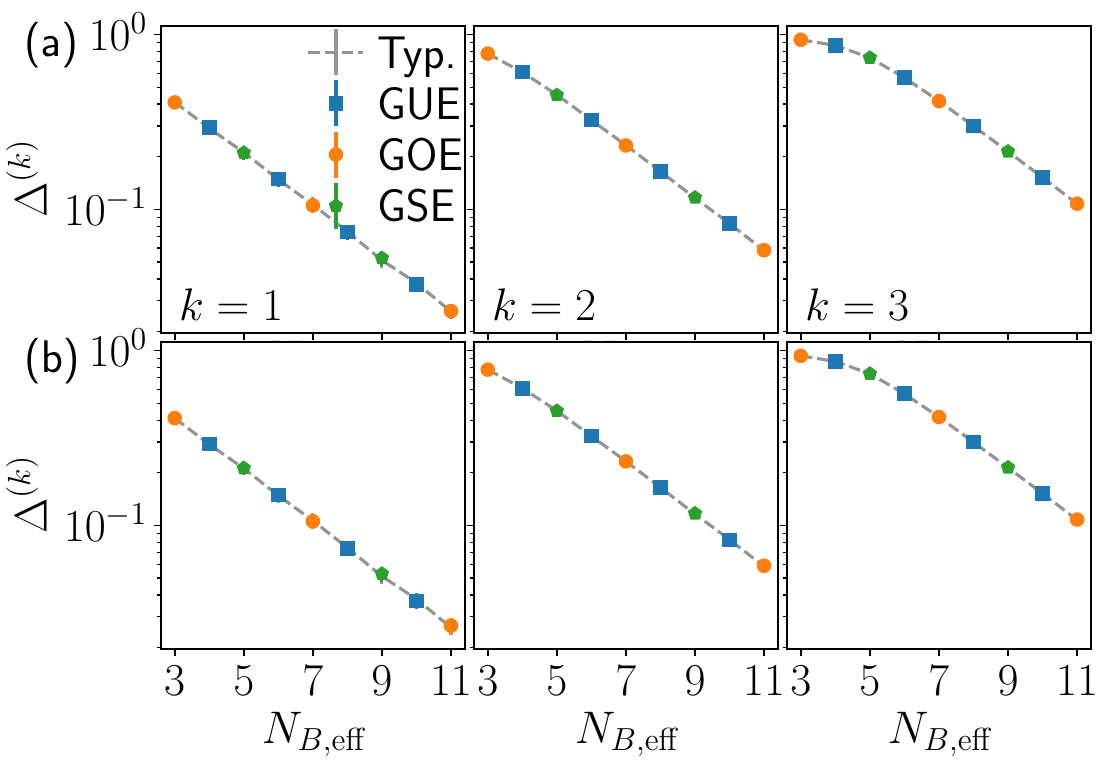}
	\caption{Late-time average of $\Delta^{(k)}$ in the SYK model with $N_A{=}4$, $\hat{Z}_P{=}(-1)^N$ and postselection for (a) a computational basis state and (b) a product state~(\ref{eq:bloch}) with random angles. 
    }
	\label{fig:SYK_ls}
\end{figure}

 Fig.~\ref{fig:SYK_full} shows the averaged late-time value of $\Delta^{(k)}$ for both types of initial states and $N_A{=}4$. To obtain a benchmark for $\Delta^{(k)}$ in a maximally chaotic state in the given Hilbert space, we also compute this value for ``typical'' states, which are simply complex random vectors where both the real and imaginary part of the wave function amplitude are drawn randomly from a normal distribution with mean 0 and variance unity. These are then pure states at infinite temperature. This benchmark value is also shown in Fig.~\ref{fig:SYK_full}, where we see it agrees closely with the results of quenches from states in Eq.~(\ref{eq:bloch}) with angles sampled randomly. By contrast, for computational basis states the exponential decay with $N_B$ quickly reaches a plateau and completely stops, suggesting that the projected ensemble never reaches the Haar ensemble.

The origin of this difference between initial states is the symmetry $\hat{Z}_P$. The computational basis states are eigenstates of this operator with eigenvalues $\alpha{=}\pm 1$, and so are the time-evolved states obtained from them. Furthermore, $\hat{Z}_P$ can be decomposed as 
\begin{equation}
    \hat{Z}_P{=}\hat{Z}_{P,A}{\otimes} \hat{Z}_{P,B}, \ \hat{Z}_{P,A}=\prod_{j=1}^{N_A} \sz_j, \ \hat{Z}_{P,B}=\hspace{-0.35cm}\prod_{j=N_A+1}^{N} \hspace{-0.25cm}\sz_j.
\end{equation}
As $\hat{Z}_{P,B}$ is diagonal in the measurement basis, the measured string $z_B$ has a definite value of $\alpha_B{=}\pm 1$ under $\hat{Z}_{P,B}$. Consequently, the corresponding state $\ket{\psi^A_{z_B}}$ must be an eigenstate of $\hat{Z}_{P,A}$ with eigenvalue $\alpha_A=\alpha/\alpha_B$. This additional constraint on the $\ket{\psi^A_{z_B}}$ prevents them from uniformly exploring $\mathcal{H}_A$ and reproducing the Haar ensemble. In contrast, states built from random angles are generically not eigenstates of $\hat{Z}_P$ and have expectation values of this operator close to 0. Thus, there is no individual constraint on the $\ket{\psi^A_{z_B}}$ and we obtain good agreement with the Haar ensemble.

In order to recover deep thermalization for all initial states, a simple solution is to restrict to the symmetry sector with an even number of $\uparrow$ spins, or alternatively $Z_P=(-1)^N$. As discussed above, this introduces correlations between the measured string $z_B$ and the possible states in $A$. We thus post-select the $z_B$ strings to only keep those with $Z_{P,B}=(-1)^{N_B}$. This implies that all $\ket{\psi^A_{z_B}}$ will obey $Z_{P,A}=(-1)^{N_A}$, effectively reducing the relevant Hilbert space of subsystem $A$. We now recover good agreement with the Haar ensemble, see Fig.~\ref{fig:SYK_ls}. Note that this reduces the dimension of the subsystem $B$ that acts as a reservoir for the states we are interested in. As such, we follow Ref.~\cite{Cotler2023} and instead of $N_B$ we keep track of its effective counterpart $N_{B,\mathrm{eff}}=\log_2(\mathcal{D}_B)$, where $\mathcal{D}_B$ is the number of postselected states in $B$ (i.e., states with with $Z_{P,B}=(-1)^{N_B}$ in the symmetry sector $Z_P=(-1)^N$). For the rest of this work, we will keep using the notation $N_{B,\mathrm{eff}}$ for consistency even when no postselection is done.

\begin{figure}[tb]
	\centering
	\includegraphics[width=\linewidth]{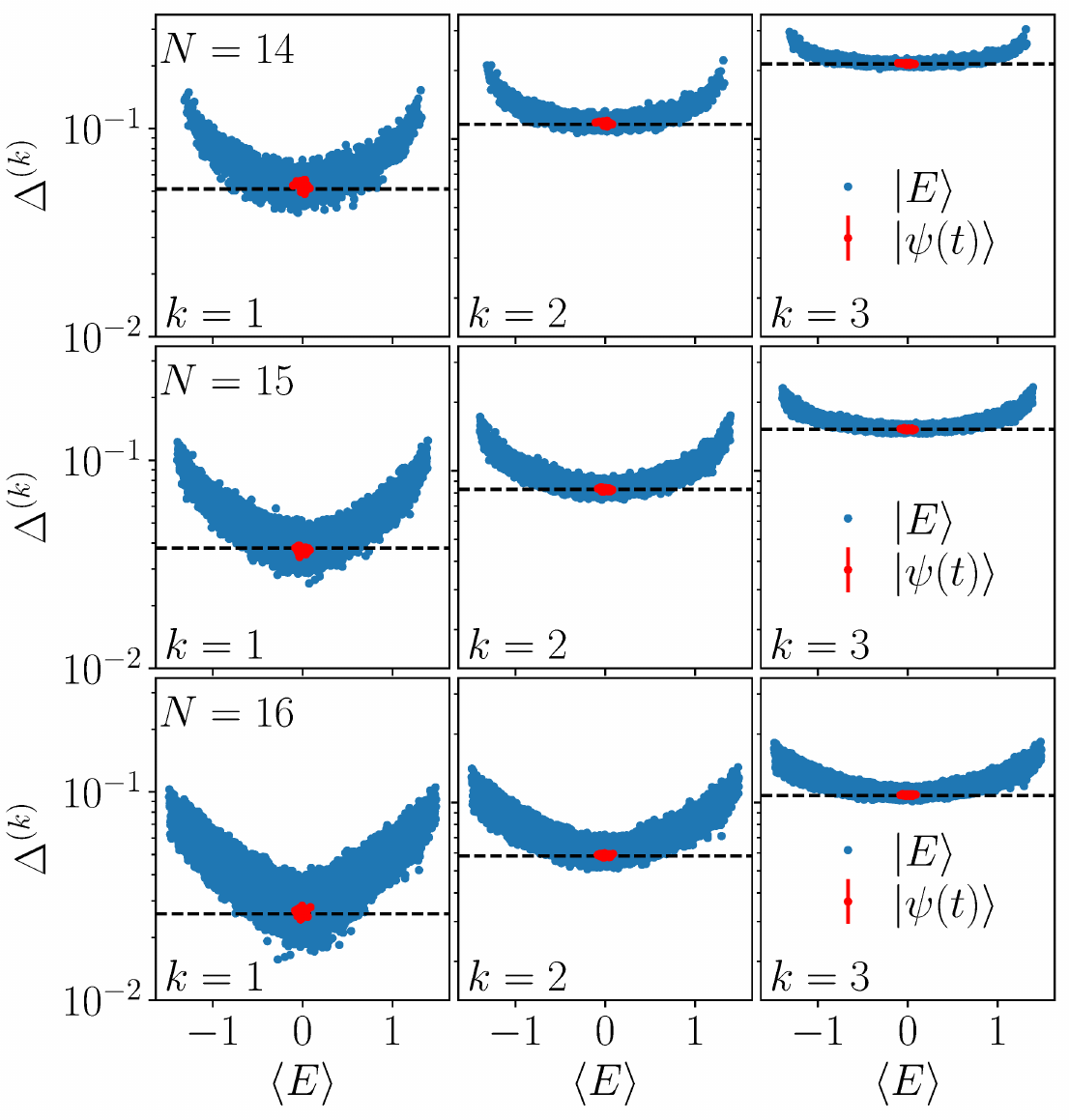}
	\caption{$\Delta^{(k)}$ for eigenstates of the SYK model for $N_A{=}4$ and $N{=}14$, $15$ and $16$, with restriction to a single symmetry sector of $\hat{Z}_P$ and postselection. The red markers indicate the long-time average for the time-evolved states shown in Fig.~\ref{fig:SYK_ls}. In all cases, there is excellent agreement between typical states, time-evolved states and eigenstates near energy $E{=}0$. This is despite the level statistics being respectively GSE ($N{=}14$), GUE ($N{=}15$) and GOE ($N{=}16$).
    }
	\label{fig:SYK_eigs}
\end{figure}

Interestingly, for the time-evolved state we see no influence of the RMT class on deep thermalization, as different particle numbers in Fig.~\ref{fig:SYK_ls} all behave identically. The computed  $\Delta^{(k)}$ for SYK eigenstates also shows no such dependence, Fig.~\ref{fig:SYK_eigs}.
Furthermore, one can also see that, for all $k$ investigated, the eigenstates at infinite temperature and time-evolved states show very close agreement, matching also the prediction for typical states. These results thus set our benchmark for deep thermalization in a fully chaotic system, which we will use for interpreting the results in the less generic models below.

\section{Time reversal}\label{sec:t_sym}

The SYK model in the previous section illustrated the effect of conventional symmetries on deep thermalization, as well as the apparent robustness of this phenomenon to the RMT class of the Hamiltonian studied. In this section, we show that for systems that have time-reversal symmetry (and so are in the GOE class), deep thermalization for eigenstates can be prevented and we formulate sufficient conditions for this to happen. 

As before, we consider a quantum model defined on a system with $N$ sites that can be divided in two subsystems with $N_A$ and $N_B$ sites, respectively. We will consider the case where the total Hilbert space is a tensor product $\mathcal{H}=\mathcal{H}_A\otimes \mathcal{H}_B$,  as this is where we expect to see deep thermalization without any additional postselection procedure.
Let us consider a Hamiltonian that is real in the measurement basis. Its eigenstates will also be real in this basis and they can be expressed as
\begin{equation}\label{eq:E}
\ket{E}=\sum_{z_B}\sqrt{p_{z_B}}\ket{\psi^A_{z_B}}\otimes \ket{z_B},
\end{equation}
with all $\ket{\psi^A_{z_B}}$ real. This implies that even if the $\ket{\psi^A_{z_B}}$ are maximally random, they will never be able to approximate the Haar ensemble, but will instead mimick the ensemble of states invariant under the application of orthogonal (instead of unitary) matrices. While this is trivial so far, we will show that the same effect can be obtained with Hamiltonians that are \emph{not} real in the measurement basis. 

Consider performing a basis change to the Hamiltonian while keeping the measurement basis the same. In particular, let us focus on the case where the change-of-basis matrix $\hat{V}$ obeys:
\begin{enumerate}
    \item{$\hat{V}$ can be decomposed into subsystems $A$ and $B$ as $\hat{V}=\hat{V}_A\otimes \hat{V}_B$, with $\hat{V}_A$ and $\hat{V}_B$ unitary;}
    \item{$\hat{V}_B$ is diagonal in the measurement basis.}
\end{enumerate} 
The corresponding projected ensemble is obtained as
\begin{equation}
\hat{V}\ket{E}=\sum_{z_B}\sqrt{p_{z_B}}e^{iv(z_B)}\left(\hat{V}_A\ket{\psi^A_{z_B}}\right)\otimes \ket{z_B},
\end{equation} 
where the term $e^{i v(z_B)}$ collects the phase change induced on each bitstring $z_B$ and adds an irrelevant overall phase term to each state in the projected ensemble. As $\hat{V}_A$ is a unitary rotation on the subsystem $A$, it cannot change the agreement between the projected ensemble and the Haar ensemble. Indeed, when comparing the two state ensembles, applying the same unitary rotation to both of them will not modify the result. So we can keep the same projected ensemble and apply $\hat{V}_A^\dagger$ to the Haar ensemble instead. But this ensemble is by definition invariant under any unitary rotation. Thus, in the end the basis change $\hat{V}$ will have no influence on $\Delta^{(k)}$ for the eigenstates. 

To construct an example where $\Delta^{(k)}$  is not the same after a basis transformation, take a change-of-basis $\hat{V}$ that instead obeys:
\begin{enumerate}
    \item{$\hat{V}$ is unitary;}
    \item{$\hat{V}$ can be decomposed into subsystems $A$ and $B$ as $\hat{V}=\sum_j \hat{V}^j_A\otimes \hat{V}^j_B$;}
    \item{All $\hat{V}^j_B$ are diagonal in the measurement basis.}
\end{enumerate}
In this case, we still have a departure from the Haar ensemble. Let us denote by $\hat{K}$ the antiunitary corresponding to complex conjugation. It is straightforward to see that $\hat{K}=\hat{K}_A\otimes \hat{K}_B$. 
As $\hat{V}\ket{E}$ is real, it follows that $\hat{K}\left(\hat{V}\ket{E}\right)=\hat{V}\ket{E}$.
We can recast both sides of this equation using Eq.~(\ref{eq:E}) to obtain
\begin{equation}\label{eq:VE}
\begin{aligned}
\hat{V}\ket{E}\!&=\!\sum_j\sum_{z_B}\sqrt{p_{z_B}}\left(\hat{V}^j_A\ket{\psi^A_{z_B}}\right)\!\otimes\!\left(\hat{V}^j_B \ket{z_B}\right)\\ &=\!\sum_{z_B}\sqrt{p_{z_B}}\left(\sum_jg_j(z_B)\hat{V}^j_A\ket{\psi^A_{z_B}}\right)\!\otimes \!\ket{z_B},\\
\hat{K}\left(\hat{V}\ket{E}\right)\!&=\!\sum_{z_B}\sqrt{p_{z_B}}\left(\hat{K}_A\!\sum_jg_j(z_B)\hat{V}^j_A\ket{\psi^A_{z_B}}\right)\!\otimes\! \ket{z_B},\\
\end{aligned}
\end{equation}
where $g_j(z_B)$ are the eigenvalues of $\ket{z_B}$ under $\hat{V}^j_B$.
As the rotation applied to $\{\ket{\psi^A_{z_B}} \}$ can now depend on $z_B$, this no longer corresponds to a global rotation of the basis in the subsystem $A$. As a consequence, $\Delta^{(k)}$ is generically not the same for $\ket{E}$ and $\hat{V}\ket{E}$. Nonetheless, as all the $\ket{z_B}$ are orthogonal, the only way to satisfy the equality between the second and last lines in Eq.~(\ref{eq:VE}) is if for \emph{all} $\ket{\psi^A_{z_B}}$
\begin{equation}
\sum_jg_j(z_B)\hat{V}^j_A\ket{\psi^A_{z_B}}=\hat{K}_A\sum_jg_j(z_B)\hat{V}^j_A\ket{\psi^A_{z_B}}
\end{equation}
holds. This puts a constraint on the individual $\ket{\psi^A_{z_B}}$, with each possible value of $z_B$ restricting the corresponding states in $A$ to a hyperplane in the Hilbert space. In specific cases, these hyperplanes can be identical or the dependence on $z_B$ can even vanish, leading to the same $\Delta^{(k)}$ as for real states.
Appendix~\ref{sec:hyperp} illustrates several different possibilities, one of which will be used in Sec.~\ref{sec:antisym} below.

\begin{figure}[tb]
	\centering
	\includegraphics[width=\linewidth]{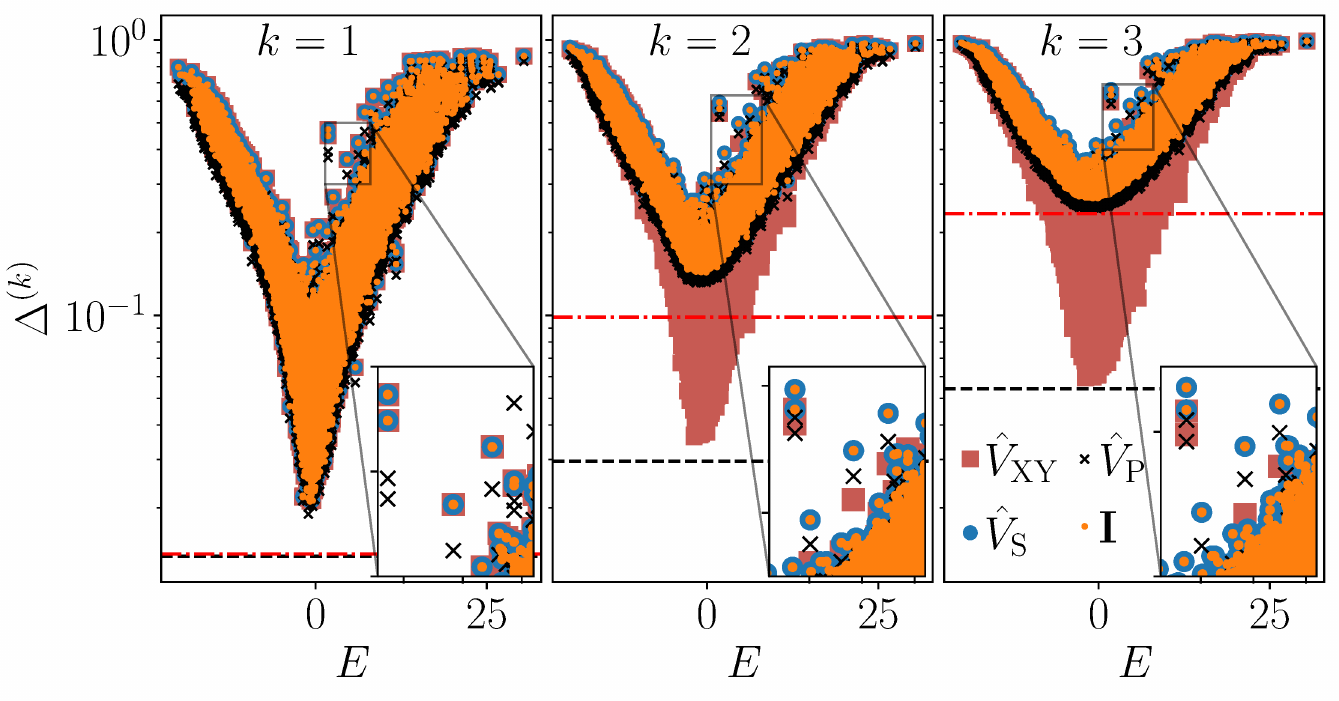}
	\caption{$\Delta^{(k)}$ for eigenstates of the quantum Ising model (\ref{eq:Ising}) for $N_A=3$ and $N=16$ after various unitary transformations generated by matrices $\hat{V}$. The dashed black (dash-dotted red) line indicates the average value for complex (real) typical states. The insets show zooms of the main panels. Only $\hat{V}_{XY}$ leads to good agreement with typical complex states.
    }
	\label{fig:Ising_eigs}
\end{figure}

\subsection{Ising model}\label{sec:ising}

The impact of basis changes can be showcased using the Ising model with both transverse and longitudinal fields:  
\begin{equation}\label{eq:Ising}
    \hat{H}_\mathrm{Ising}=\sum_{j=1}^{N-1} \sz_j\sz_{j+1}+h\sum_{j=1}^{N}\sz_j+g\sum_{j=1}^{N}\sx_j,
\end{equation}
where we use the parameters $h=(1+\sqrt{5})/4$ and $g=(\sqrt{5}+5)/8$ that were found in Ref.~\cite{KimHuse2013} to lead to strongly chaotic dynamics. In order to explore the different cases, we define various Hamiltonians obtained through the change of basis $\hat{H}^\prime=\hat{V}\hat{H}_\mathrm{Ising}\hat{V}^\dagger$. We consider three different change-of-basis matrices $\hat{V}$: 
\begin{align}
   \hat{V}_{S}&=\exp\left(-\frac{i\pi}{8}\sum_j \sz_j \right), 
   \\
   \hat{V}_{P}&=\exp\left(-\frac{i\pi}{8}\prod_j \sz_j \right),\\
  \hat{V}_\mathrm{XY} &= \exp\left(\frac{i\pi}{4}\sum_j \sy_j \right)\exp\left(\frac{i\pi}{4}\sum_j \sz_j \right).
\end{align}
A few remarks are in order. While it is clear that both  $\hat{V}_S$ and $\hat{V}_\mathrm{XY}$ can be decomposed as $\hat{V}_A\otimes \hat{V}_B$, $\hat{V}_B$ is only diagonal for $\hat{V}_S$. $\hat{V}_{P}$ cannot be decomposed in the same way, but it can be rewritten as $\hat{V}_{P}=\cos(\pi/8)\mathbb{1}-i\sin(\pi/8)\prod_j \sz_j$. This means it corresponds to the case  $\hat{V}=\sum_j \hat{V}^j_A\otimes \hat{V}^j_B$ with all $\hat{V}^j_B$ diagonal. Finally, we note that performing the transformation $\hat{V}_\mathrm{XY}\hat{H}_\mathrm{Ising}\hat{V}_\mathrm{XY}^\dagger$ leads to the same Ising Hamiltonian considered in Ref.~\cite{Cotler2023}, for which the eigenstates at infinite temperature display good agreement with the Haar ensemble. 

The values of $\Delta^{(k)}$ with $k=1$ to $3$ for  different bases are shown in  Fig.~\ref{fig:Ising_eigs}. For $k=1$, the eigenstates of the Hamiltonians generated by $\mathds{I}$, $\hat{V}_{S}$ and $\hat{V}_\mathrm{XY}$ show the \emph{exact} same values of $\Delta^{(k)}$. However, this is no longer the case for $k>1$, where $\mathds{I}$ and $\hat{V}_{S}$ are still identical but greatly differ from $\hat{V}_\mathrm{XY}$ due to the states in the projected ensemble being real. This illustrates the difference between the physical case of $k=1$, which is directly linked to the expectation values of observables in subsystem $A$ and thus to ETH, and $k>1$. Meanwhile, for $\hat{V}_{P}$ we see that $\Delta^{(k)}$ is different from the other cases, but the constraint on the states in the projected ensemble still prevents deep thermalization even at infinite temperature. In Appendix~\ref{sec:hyperp}, the dependence of deep thermalization on the angle used in $\hat{V}_P$ is also explored. 
While these results on deep thermalization of eigenstates are not directly relevant for experiments since preparing high-energy eigenstates is a difficult task, we will show in Sec.~\ref{sec:antisym} that the same kind of mechanism can also prevent a time-evolved state from thermalizing.

\section{Quantum East model}\label{sec:PX}

Now that we have understood deep thermalization in the SYK model and the pitfalls associated with time reversal and basis transformations, we move on to our first kinetically-constrained model: the quantum East model~\cite{vanHorssen2015,Pancotti2020}. The model is inspired by classical structural glasses and features two different phases: one that has slow thermalization and localized eigenstates, and the other where local observables appear to thermalize from the point of view of ETH \cite{Riccardo_East_PRX_2022}. We focus on the latter phase to see if deep thermalization can detect any hidden non-ergodicity in this regime. 

The quantum East model is defined on a 1D lattice with spin-1/2 degrees of freedom and the Hamiltonian
\begin{equation}\label{eq:East_OBC}
    \hat{H}_\mathrm{qEast}=\sx_1+\sum_{j=2}^N \Pz_{j-1}\sx_j,
\end{equation}
where $\Pz_j =(1-\sz_j)/2$ is a local projector on the spin-down state at site $j$. The projector enforces the following constraint: a spin can only flip if its left neighbor is in the $\downarrow$ state, previously illustrated in Fig.~\ref{fig:illustration_PE}(b).
We will first consider open boundary conditions (OBCs) in order to get rid of lattice momentum as a conserved quantity. In this case, the first site does not have a neighbor to the left, thus the Hamiltonian term is only $\hat\sigma_1^x$, as indicated in Eq.~(\ref{eq:East_OBC}). Note that $\hat{H}_\mathrm{qEast}$ commutes with the local operator $\sx_N$. However, as the latter is fully off-diagonal in the measurement basis, it should not introduce any correlation between subsystems $A$ and $B$. As such we do not limit our study to a single sector of it. In addition, $\hat{H}_\mathrm{qEast}$ is real and anticommutes with $\hat{Z}=\prod_{j=1}^N \sz_j$. Since this is not a conserved quantity, we take no additional step in the computation of $ \Delta^{(k)}$ because of it.

\begin{figure}[tb]
	\centering
	\includegraphics[width=\linewidth]{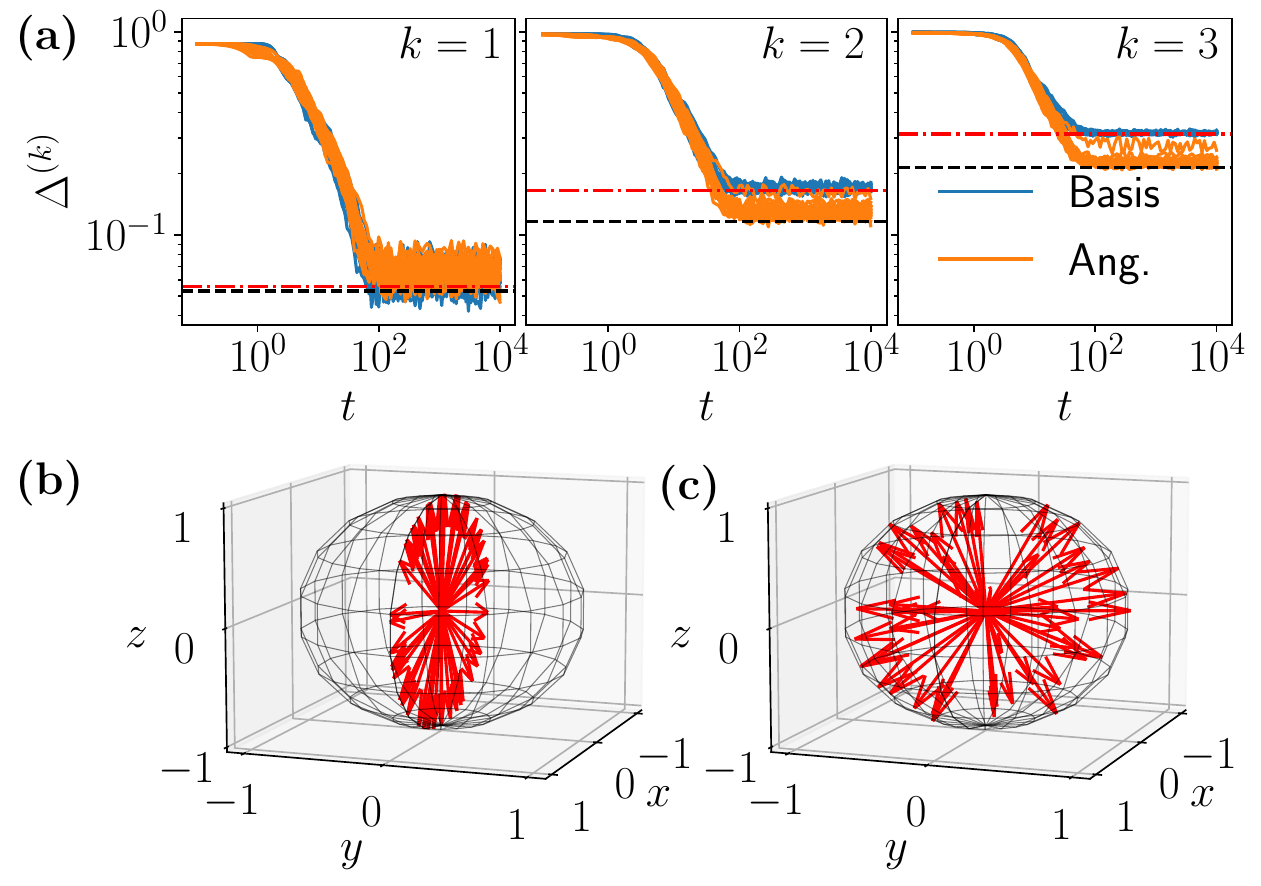}
	\caption{Deep thermalization in the quantum East model for $N=12$. (a) $\Delta^{(k)}$ with $N_A=3$ for two different types of initial states at infinite temperature. The dashed black lines (dash-dotted red lines) indicate the average $\Delta^{(k)}$ for (real) states. (b)-(c) For $N_A=1$, we plot 50 states from the projected ensemble of a time-evolved (b) basis state and (c) a state with random angles at $t=10000$. For the former, all vectors are in the $YZ$ plane while for the latter they explore all sectors of the Bloch sphere.
    }
	\label{fig:PX_Bloch}
\end{figure}

However, when performing quenches from random basis states and states built from random angles (\ref{eq:bloch}), we see a stark difference between the two for $k>1$, Fig.~\ref{fig:PX_Bloch}(a). For the initial states (\ref{eq:bloch}), here we set the $\phi$ angle of the last site to $\pi/2$ to always be equally split between the two symmetry sectors of $\sx_N$. While the random-angle initial states deep thermalize as expected, Fig.~\ref{fig:PX_Bloch}(a) shows that random basis states display the same value of $\Delta^{(k)}$ as \emph{real} typical states. The source of this difference is not found in symmetries this time, but rather in the anticommuting operator $\hat{Z}$.
Indeed, all the basis states considered are eigenstates of $\hat{Z}$ with eigenvalue $\alpha=\pm 1$. As they are also real, they are eigenstates of $\hat{K}\hat{Z}$, where $\hat{K}$ is the complex conjugation operator. Finally, as the Hamiltonian is real, this allows to write
\begin{equation}\label{eq:KZ}
    \hat{K}\hat{Z}e^{-i\hat{H}t}\ket{\psi}{=}\hat{K}e^{i\hat{H}t}\hat{Z}\ket{\psi}{=}e^{-i\hat{H}t}\hat{K}\hat{Z}\ket{\psi}{=}\alpha \ket{\psi (t)},
\end{equation}
implying that the time-evolved states are eigenstates of $\hat{K}\hat{Z}$.
The $\ket{z_B}$ states are eigenstates of $\hat{K}_B$ with eigenvalues 1 and eigenstates of $\hat{Z}$ with eigenvalue $\alpha_{z_B}=\pm 1$. This allows us to rewrite, using Eq.~(\ref{eq:KZ}), 
\begin{equation}\label{eq:global}
\begin{aligned}
   \alpha\ket{\psi(t)}&=\alpha \sum_{z_B}\sqrt{p_{z_B}}\ket{\psi^A_{z_B}}\otimes \ket{z_B}\\ &=\hat{K}\hat{Z}\ket{\psi(t)}=\hat{K}\hat{Z}\sum_{z_B}\sqrt{p_{z_B}}\ket{\psi^A_{z_B}}\otimes \ket{z_B}. 
\end{aligned}
\end{equation}
Since $\hat{Z}$ can be rewritten as $\hat{Z}=\hat{Z}_A\otimes \hat{Z}_B$, and the same is trivially true for $\hat{K}$, we have: 
\begin{equation}\label{eq:global2}
\begin{aligned}
   \alpha\ket{\psi(t)} &=\sum_{z_B}\sqrt{p_{z_B}}\left(\hat{K}_A\hat{Z}_A\ket{\psi^A_{z_B}}\right)\otimes      \left(\hat{K}_B\hat{Z}_B\ket{z_B}\right) \\
&=\sum_{z_B}\sqrt{p_{z_B}}\alpha_{z_B}\left(\hat{K}_A\hat{Z}_A\ket{\psi^A_{z_B}}\right)\otimes \ket{z_B}.
\end{aligned}
\end{equation}
As all $\ket{z_B}$ are orthogonal, it must hold that 
\begin{equation}\label{eq:local}
\hat{K}_A\hat{Z}_A\ket{\psi^A_{z_B}}=\frac{\alpha}{\alpha_{z_B}}\ket{\psi^A_{z_B}}, \quad \forall \ z_B.
\end{equation}
As in the case of real eigenstates, this puts a constraint on the individual $\ket{\psi^A_{z_B}}$. 

\begin{figure}[tb]
	\centering
	\includegraphics[width=\linewidth]{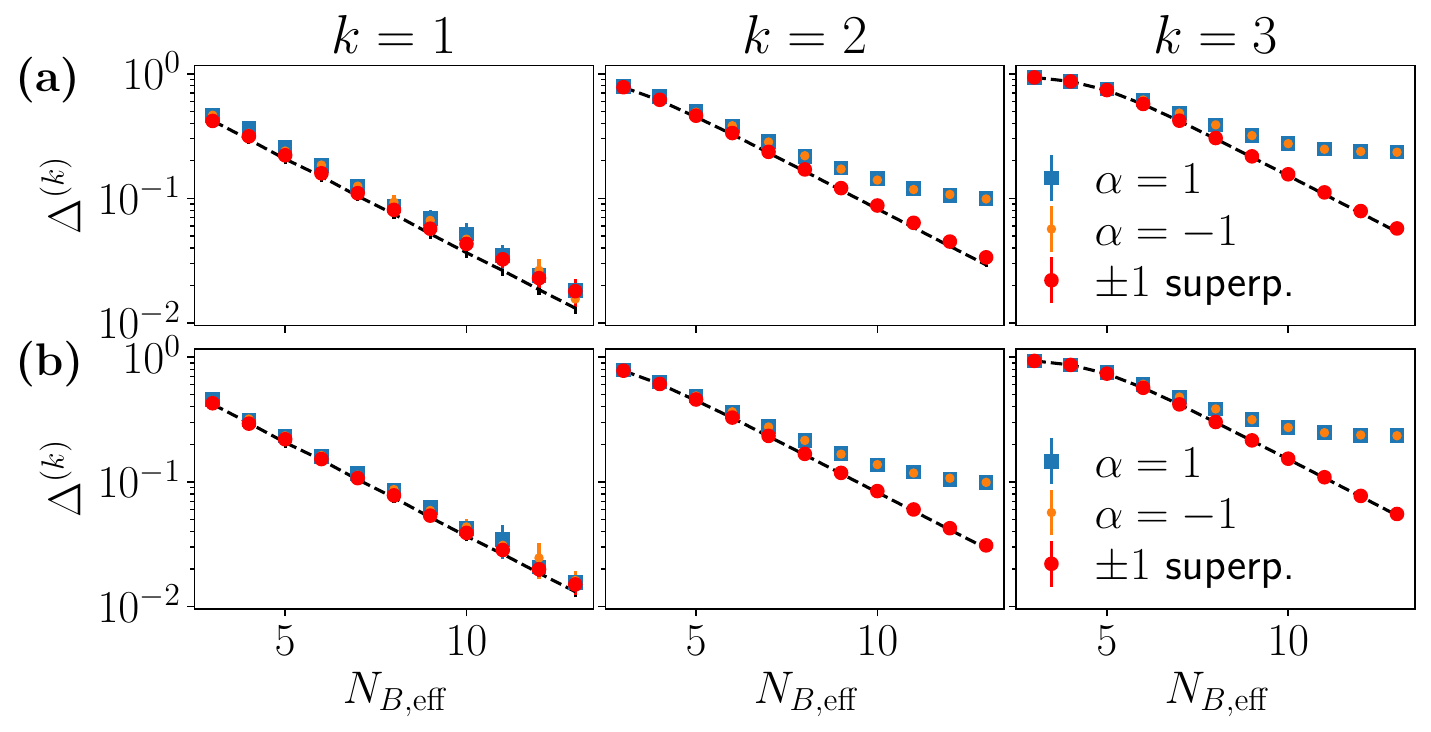}
	\caption{Long-time average of $\Delta^{(k)}$ in the quantum East model with $N_A=3$ and various system sizes for (a) OBCs and (b) PBCs. Both types of boundary conditions show very similar results. The effect of the anticommuting operator $\hat Z$ is clearly visible when comparing initial states. 
    }
	\label{fig:PX_quench}
\end{figure}

To see the consequences of the constraint (\ref{eq:local}), let us consider the case of a single qubit in the projected ensemble with $\alpha{=}1$ and $N_A{=}1$, leading to $\hat{Z}_A{=}\sz_1$.
If we parameterize the states in the projected ensemble as $\ket{\theta,\phi}=\cos(\theta/2)\ket{\downarrow}+e^{i\phi}\sin(\theta/2)\ket{\uparrow}$, we get 
\begin{equation}
        \hat{K}_1 \sz_1\ket{\theta,\phi}=-\cos(\theta/2)\ket{\downarrow}+e^{-i\phi}\sin(\theta/2)\ket{\uparrow}.
\end{equation}
We end up with $\theta$ free, but $\phi=\pm \pi/2$. If we plot the $\ket{\psi^A_{z_B}}$ on the Bloch sphere, they will thus only lie in the $YZ$ plane where $\phi=\pm \pi/2$. This is confirmed in the numerical simulation in Fig.~\ref{fig:PX_Bloch}(b). Confining the projected states to a hyperplane is enough to reproduce the mean (first moment) of the Haar ensemble, but it is inadequate to reproduce the higher moments, which lies at the root of the disparity between $k=1$ and $k>1$. If, instead, the initial state is not an eigenstate of $\hat{Z}$, then Eq.~(\ref{eq:global}) does not put an individual constraint on each state in the projected ensemble. As such, states in the projected ensemble are unconstrained (see Fig.~\ref{fig:PX_Bloch}) and we can expect deep thermalization for all $k$. To test this, we compare the late-time average of $\Delta^{(k)}$ of basis states with $\alpha{=}1$, $\alpha{=}-1$, as well as symmetric superpositions $(\ket{\alpha{=}+1}+\ket{\alpha{=}{-1}})/\sqrt{2}$. This is shown in Fig.~\ref{fig:PX_quench} (a), which reveals clear signatures of deep thermalization only for the latter states.

\begin{figure}[t!]
	\centering
	\includegraphics[width=\linewidth]{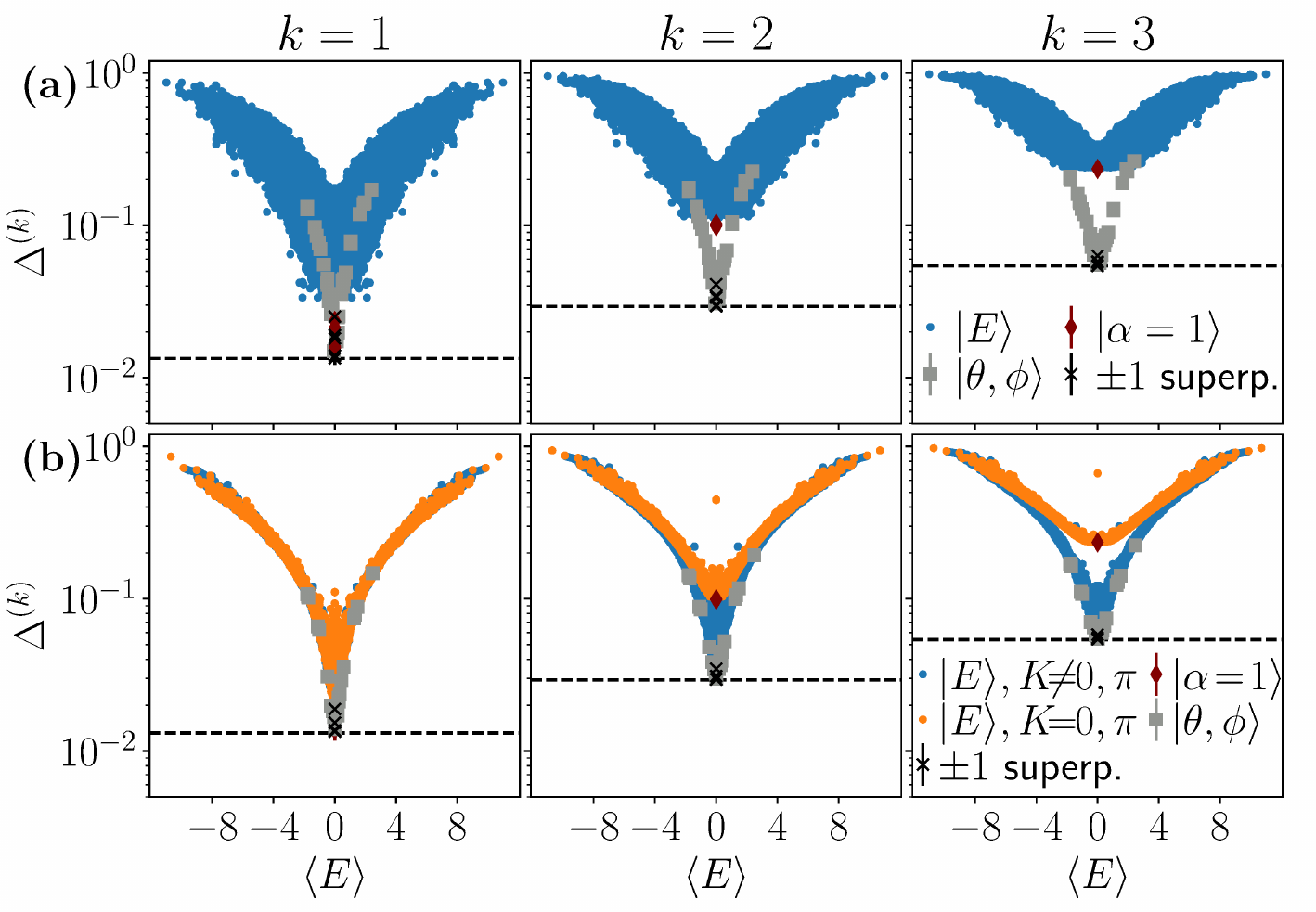}
	\caption{$\Delta^{(k)}$ for the eigenstates in the quantum East model with $N_A=3$ and $N=14$ with (a) OBCs and (b) PBCs. The effect of the eigenstates being real is clearly visible for all $\ket{E}$ with OBCs, and for momenta $K=0$ and $K=\pi$ for PBCs. 
    }
	\label{fig:PX_eigs}
\end{figure}

In order to verify that the observed deviation from deep thermalization is not due to the conserved charge $\sx_N$, we can compare our results with the same model with periodic boundary conditions (PBCs). The Hamiltonian is the same as in Eq.~(\ref{eq:East_OBC}), except for the first term that becomes $\hat{P}_{N}\sx_1$. Now, $\sx_N$ no longer commutes with the Hamiltonian, however, we have a conservation of the lattice momentum. Note that with PBCs the state $\ket{{\uparrow}{\uparrow}\cdots {\uparrow}}$ is also disconnected from the rest of the Hilbert space, and we explicitly discard it. The results in Fig.~\ref{fig:PX_quench} (b) show that OBCs and PBCs give extremely similar results for $\Delta^{(k)}$ at late times. As they are also essentially identical to those of typical states, this shows that neither  $\sx_N$ nor lattice momentum affect deep thermalization. However, this is no longer true for eigenstates, as seen in Fig.~\ref{fig:PX_eigs}. Momentum-resolved eigenstates are no longer real, except in the $K=0$ and $K=\pi$ sectors. Thus, we can see a clear difference between these sectors and and the rest. Interestingly, we also see a much narrower distribution of  $\Delta^{(k)}$, even in real sectors.

In summary, we find that the quantum East model shows clear signatures of deep thermalization for both eigenstates and time-evolved states, once the relevant symmetries and anticommuting operators are properly accounted for. We address the role of the anticommuting operators and the conditions for deep thermalization more generally in the subsequent section. 

\section{Anticommuting operators and time reversal}\label{sec:antisym}

While the Hamiltonian having time-reversal symmetry can change the ensemble to which the eigenstates deep-thermalize to, this is generally not the case for time-evolved state. Indeed, even if the initial state and Hamiltonian are real in the measurement basis, this will not be the case of the time-evolved state. 
However, for a Hamiltonian that is real in the canonical basis, we can recover the same constraints for the projected ensemble if there exists an observable $\hat{Z}$ that: 
\begin{enumerate}
    \item {anticommutes with the Hamiltonian,  $\{ \hat Z, \hat H\} = 0$;}
    \item{Can be decomposed into subsystems $A$ and $B$ as $\hat{Z}=\hat{Z}_A\otimes \hat{Z}_B$ with $\hat{Z}_A$ and $\hat{Z}_B$ hermitian;}
    \item{$\hat{Z}_B$ is diagonal in the measurement basis;}
    \item{$\hat{Z}$ only has eigenvalues $+1$ and $-1$ and squares to the identity.}
\end{enumerate}
Note that these conditions imply that $\hat{Z}_A$ and $\hat{Z}_B$ both have eigenvalues $\pm 1$ and square to the identity. 

Let us assume that our initial state $\ket{\psi}$ is real (up to a global phase) in the measurement basis and an eigenstate of $\hat{Z}$ with eigenvalue $\alpha=\pm 1$. Then one can perform a change of basis using 
\begin{equation}
   \hat{V}=\exp\left[i\frac{\pi}{4} (\hat{Z}-\alpha{\mathbb{1}})\right]=\frac{1-\alpha i}{2}\left({\mathbb{1}}+i\hat{Z}\right).
\end{equation}
The resulting time-evolved state will then be 
\begin{equation}
\begin{aligned}
\hat{V}\ket{\psi(t)}&=\frac{1-i\alpha}{2}\left({\mathbb{1}}+i\hat{Z}\right)e^{-i\hat{H}t}\ket{\psi}\\
&=\frac{1-i\alpha}{2}\left(e^{-i\hat{H}t}\ket{\psi}+ie^{i\hat{H}t}\hat{Z}\ket{\psi}\right) \\
&=\left(\frac{1-i\alpha }{2}e^{-i\hat{H}t}+\frac{1+i\alpha}{2} e^{i\hat{H}t}\right)\ket{\psi}. \\
\end{aligned}
\end{equation}
The expression in the last line is clearly real, and as such cannot lead to thermalization to the Haar ensemble.

Now we can recognize that, if $\hat{Z}=\hat{Z}_A\otimes \hat{Z}_B$, then $\hat{V}$ cannot be decomposed in the same way unless $\hat{Z}_B$ acts as the identity on $B$. However, due to the specific form of $\hat{V}$, we can show that we still get convergence towards an ensemble similar to that of real states. Let us denote by $\hat{R}$ the real state such that $\ket{\psi(t)}=\hat{V}^\dagger\ket{R}$. We can then write
\begin{equation}\label{eq:rot_ZA}
\begin{aligned}
   \ket{\psi(t)}&=\exp\left(-i\frac{\pi}{4} (\hat{Z}-\alpha{\mathbb{1}})\right)\ket{R} \\   
   &=\frac{1{+}\alpha i}{2}\sum_{z_B}\sqrt{p_{z_B}}\left({\mathbb{1}}-i\alpha_{z_B}\hat{Z}_A \right)\ket{\tilde{\psi}^A_{z_B}}\!\otimes \! \ket{z_B}\\
   &=\frac{1{+}\alpha i}{\sqrt{2}}\sum_{z_B}\sqrt{p_{z_B}}\exp\left(-i\alpha_{z_B}\frac{\pi}{4}\hat{Z}_A\right)\ket{\tilde{\psi}^A_{z_B}}\!\otimes \!\ket{z_B},
\end{aligned}
\end{equation}
where the $\ket{\tilde{\psi}^A_{z_B}}$ denote the states in the projected ensemble of $\ket{R}$. These states all live in the real hyperplane of the Hilbert space of $A$. The action of $\exp\left(-i\alpha_{z_B}\frac{\pi}{4}\hat{Z}_A\right)$ rotates this hyperplane along the axis specified by $\hat{Z}_A$. As the angle of rotation is $\pi/4$ and the eigenvalues of $\hat{Z}_A$ are $\pm 1$, having $\alpha_{z_B}$ equal to plus or minus one ends up in the same hyperplane (see Appendix~\ref{sec:hyperp} for an example). This means that the $\ket{\psi^A_{z_B}}$ also live in a hyperplane of the same dimensionality as the one of real states. As a consequence, $\Delta^{(k)}$ will have the same lower bound as for real states. Still, we emphasize here that as the direction of the rotation depends on $\alpha_{z_B}$, the $\Delta^{(k)}$ for the ensemble $\{\ket{\tilde{\psi}^A_{z_B}}\}$ (corresponding to $\ket{R}$) and $\{\ket{\psi^A_{z_B}}\}$ (corresponding to $\ket{\psi(t)}$) will not be identical but will only approach the same value as $N_B\to \infty$.

Some of the stated conditions can actually be relaxed. Indeed, the requirement that both the initial state and Hamiltonian are real can be changed by requiring that there exists a change of basis $\hat{U}$ that satisfies $\hat{U}=\hat{U}_A\otimes \hat{U}_B$ with $\hat{U}_B$ diagonal in the measurement basis and such that $\hat{U}\hat{H}\hat{U}^\dagger$ and $\hat{U}\ket{\psi}$ are real (up to a global phase). One can then use $\hat{U}\hat{V}$ to turn $\ket{\psi(t)}$ into a real vector. 

\section{PXP model}\label{sec:PXP}

Finally, now that we have understood the different conditions that can prevent deep thermalization due to symmetries and anticommuting operators, we consider a model in which the constraint is stronger than in the East model and actually splits the Hilbert space into exponentially many, dynamically disconnected sectors. The model we study is the PXP model, which describes a 1D chain of Rydberg atoms~\cite{FendleySachdev,Lesanovsky2012, Bernien2017}:
\begin{equation}\label{eq:PXP}
\hat{H}_\mathrm{PXP}=\sx_1\Pz_{2}+\Pz_{N-1}\sx_N+\sum_{j=2}^{N-1} \Pz_{j-1}\sx_j\Pz_{j+1}.
\end{equation}
Recall that $\Pz_j=(1-\sz_j)/2$ is a projector on the $\downarrow$ spin state on site $j$. In this model, the projectors physically originate from van der Waals interactions between Rydberg atoms,  preventing the creation or destruction of neighboring $\uparrow$ spins, which is also known as the Rydberg blockade regime~\cite{Labuhn2016}. In our study, we restrict to the largest connected sector, the one without any configurations that contain neighboring $\cdots \uparrow\uparrow \cdots$. In Eq.~(\ref{eq:PXP}) we have assumed OBCs and separated out the boundary terms where the projectors falling outside the boundaries of the chain have been replaced by identity operators.

\begin{figure}[tb]
	\centering
	\includegraphics[width=\linewidth]{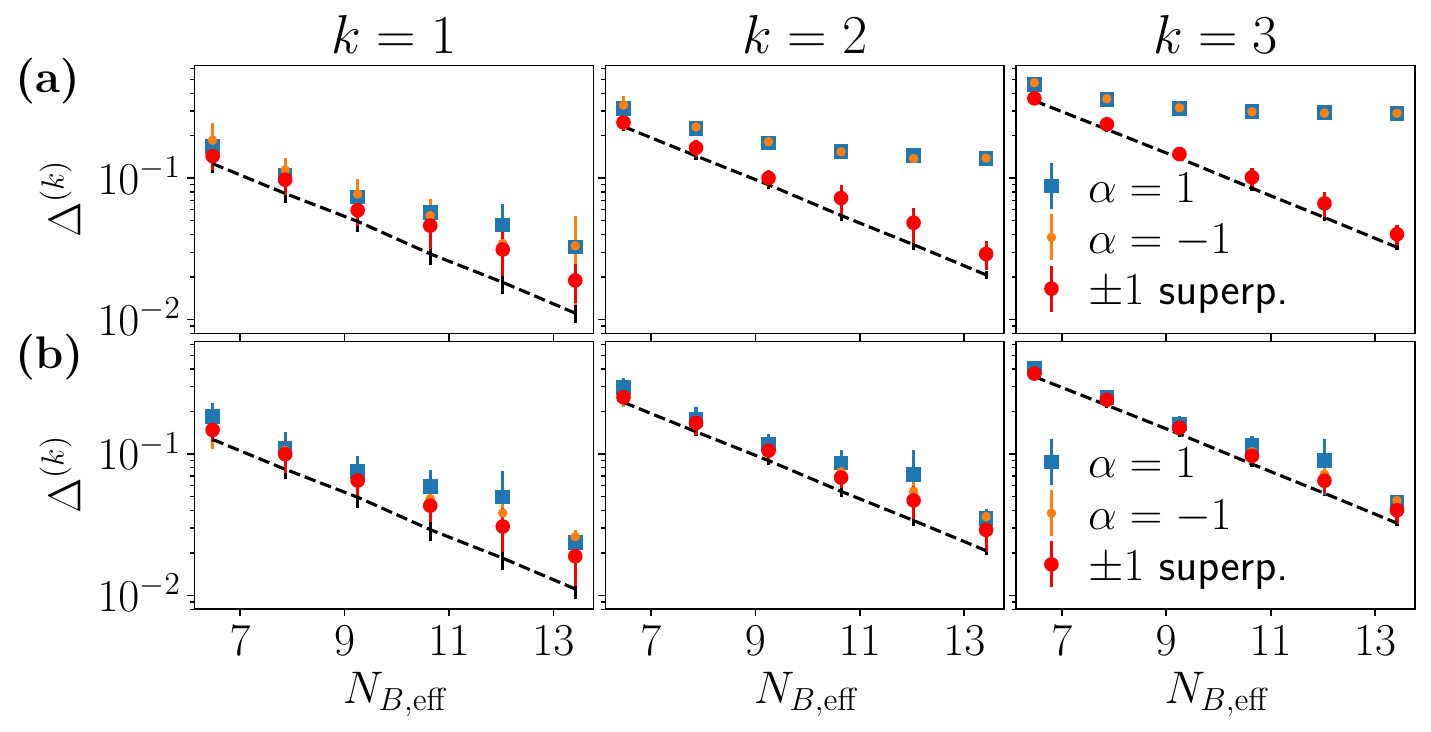}
	\caption{Late-time average of $\Delta^{(k)}$ in the PXP model with $N_A{=}3$ and various system sizes with (a) $\mu{=}0$ and (b) $\mu{=}0.05$. While for $k{=}1$ there is very little difference between the two, the contrast is strong for $k{>}1$ in large systems. 
    }
	\label{fig:PXP_quench}
\end{figure}

The PXP model is chaotic but it hosts non-thermal eigenstates known as quantum many-body scars (QMBSs) \cite{Bernien2017,Turner2018a}. Signatures of QMBSs are visible in global quenches from initial states that have high-overlap with these non-thermalizing QMBS eigenstates. Such quenches have been shown to lead to anomalous dynamics with long-lived coherent oscillations and slow thermalization, despite the model overall displaying chaotic level statistics~\cite{Turner2018b}. In particular, the states $\ket{\mathbb{Z}_2}=\ket{\uparrow \downarrow \uparrow \downarrow \cdots \uparrow \downarrow}$ and $\ket{\mathbb{Z}_3}=\ket{\uparrow \downarrow \downarrow \uparrow \downarrow \downarrow \cdots \uparrow  \downarrow\downarrow}$ show persistent quantum revivals. The origin of these revivals has been understood within a semiclassical approximation~\cite{wenwei18TDVPscar, Michailidis2019, turner2020correspondence}, which established a parallel between this many-body phenomenon and quantum scars of a single particle inside a stadium billiard~\cite{Heller84}. On the other hand, states $\ket{\mathbb{Z}_4}=\ket{\uparrow \downarrow \downarrow \downarrow \uparrow \downarrow \downarrow \downarrow \cdots \uparrow \downarrow \downarrow \downarrow}$ and $\ket{\mathbb{Z}_0}=\ket{ \downarrow \downarrow \cdots \downarrow \downarrow}$ show fast thermalization, consistent with a generic chaotic model. Note that, since the PXP Hamiltonian (\ref{eq:PXP}) is purely off-diagonal in the computational basis, all these initial states are effectively at infinite temperature. Therefore, according to strong ETH~\cite{Ueda2020}, they are expected to give rise to similar dynamics. This type of ETH violation therefore represents a form of ``weak'' ergodicity breaking~\cite{Serbyn2021}.

\begin{figure}[bt]
	\centering
	\includegraphics[width=\linewidth]{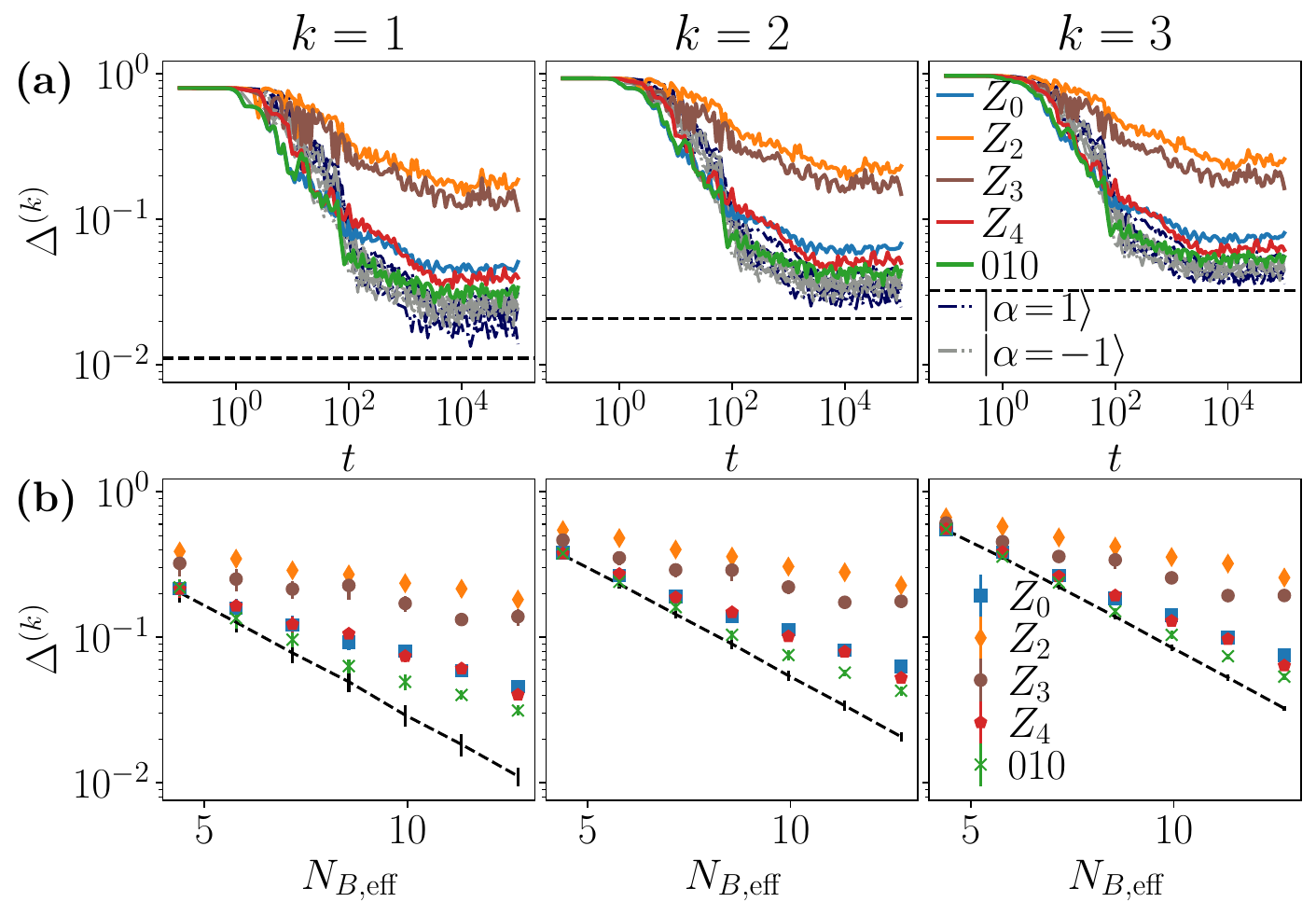}
	\caption{$\Delta^{(k)}$ for time-evolved states in the PXP model with $N_A=3$ and $\mu=0.05$. We contrast the states $\ket{\mathbb{Z}_2}$ and $\ket{\mathbb{Z}_3}$, which give rise to QMBS dynamics, with $\ket{\mathbb{Z}_0}$ and $\ket{\mathbb{Z}_4}$ that do not exhibit scarring revivals.  We also consider the ``010'' state which contains a single $\uparrow$ spin in the middle of $\ket{\mathbb{Z}_0}$.
    (a) Time series data for $N=22$. (b) Late-time average for several system sizes. The strong influence on initial states is clearly visible, but all states still show a decrease of $\Delta^{(k)}$ with $N_B$. However, even generic states and thermalizing states like $\ket{010}$ show a very slow decay of $\Delta^{(k)}$, with the plateau only reached at times of order $10^3$.
    }
	\label{fig:PXP_init}
\end{figure}

We now probe deep thermalization in the PXP model. 
Due to the Rydberg blockade, as explained in Sec.~\ref{sec:syms}, we perform postselection on the $z_B$ and only keep the strings where the first spin in $B$ (the one next to subsystem $A$) is $\downarrow$. The same postselection was used in Ref.~\cite{Choi_nature_2023}, which studied the full Rydberg model where the blockade is imperfect, i.e., it is enforced as a finite energetic penalty rather than as hard constraint like in our Eq.~(\ref{eq:PXP}).
Moreover, in light of Sec.~\ref{sec:antisym}, we must keep in mind that the PXP model in Eq.~(\ref{eq:PXP}) anticommutes with $\hat{Z}=\prod_{j=1}^N \sz_j$, similar to the East model. This operator has also been referred to as ``particle-hole transformation" in this context and it was shown to give rise to a large subspace of exact zero energy states~\cite{Turner2018a, Iadecola2018, lin2018exact}.

From our discussion above, we expect a strong dependence of $\Delta^{(k)}$ on the initial state and not just on its energy.  However, all the special states $\ket{\mathbb{Z}_n}$ are eigenstates of $\hat{Z}$ with $\alpha{=}\pm 1$, hence we know they will never exhibit deep thermalization. We can remedy this by adding a small chemical potential in the form of 
\begin{equation}
\hat{H}=\hat{H}_\mathrm{PXP}+\hat{H}_{\mu}=\hat{H}_\mathrm{PXP}+\mu \sum_{j=1}^N \hat{n}_j, 
\end{equation}
where $\hat{n}_j{=}1{-}\Pz_j$ takes the value 1 if $j$th spin points $\uparrow$ (and 0 otherwise). As $\hat{Z}$ anticommutes with $\hat{H}_\mathrm{PXP}$ but commutes with $\hat{H}_{\mu}$, it neither commutes nor anticommutes with $\hat{H}$ as long as $\mu{\neq} 0$. We thus set $\mu{=}0.05$, which causes all states to deep-thermalize but otherwise does not drastically affect the dynamics, as shown on Fig.~\ref{fig:PXP_quench}. In particular, the initial states considered remain close to infinite temperature and QMBS revivals are present when quenching from $\ket{\mathbb{Z}_2}$ and $\ket{\mathbb{Z}_3}$ states. Fig.~\ref{fig:PXP_quench} shows that even a small $\mu$ value is sufficient to make a generic state (even those with $\alpha{=}\pm 1$) deep-thermalize at late times. It also shows that while the agreement with typical states is not as good as for the models considered previously, we recover an exponential decay with $N_B$.

\begin{figure}[tb]
	\centering
	\includegraphics[width=\linewidth]{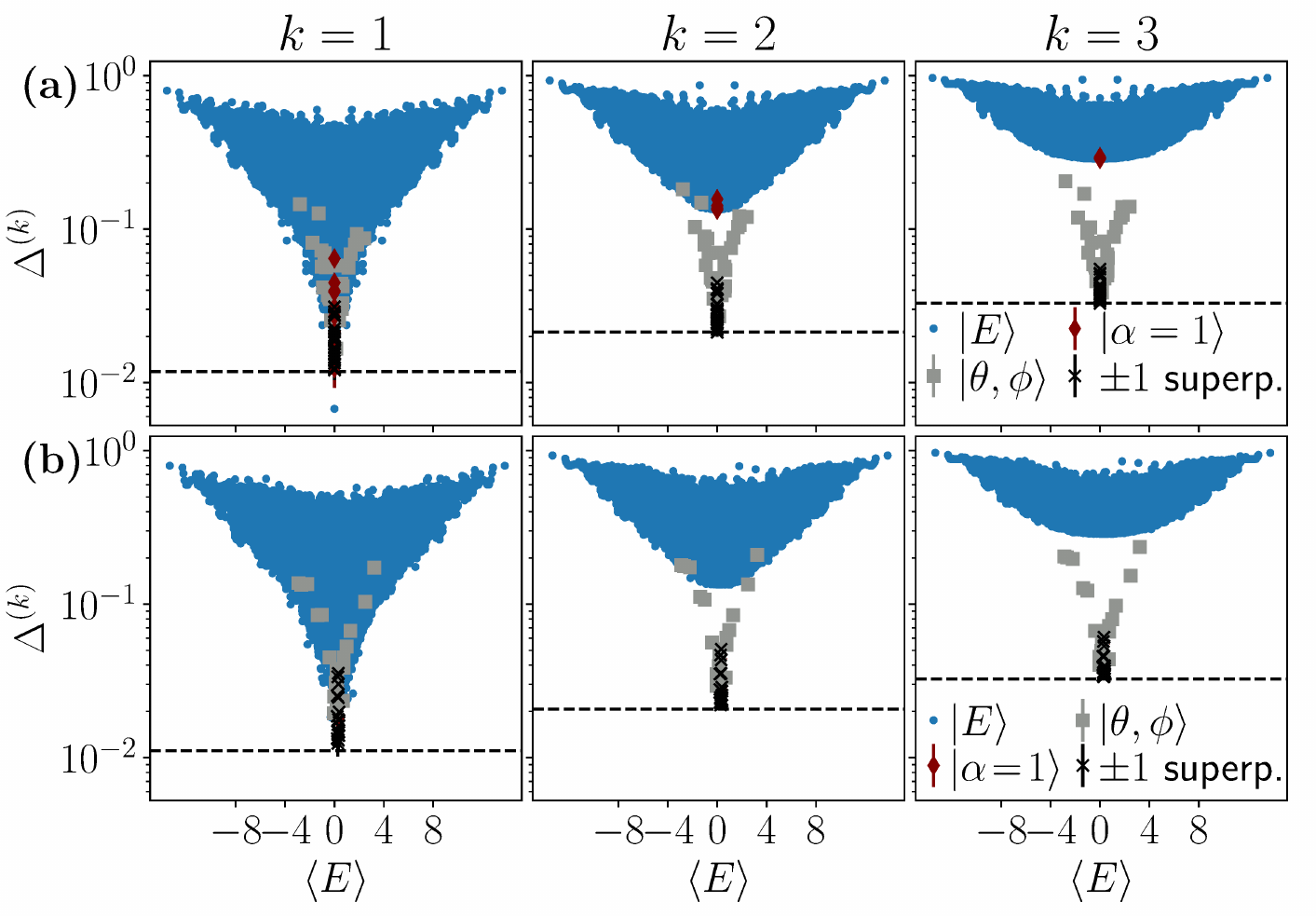}
	\caption{$\Delta^{(k)}$  for the eigenstates of the PXP model with $N_A{=}3$ and $N{=}22$ with (a) $\mu{=}0$ and (b) $\mu{=}0.05$. The late-time averages of time-evolved states are also shown. The effect of eigenstates being real is clearly visible for all $\ket{E}$, while $\mu$ has no strong influence for the eigenstates. Near $E=0$, the range of values of $\Delta^{(k)}$ is much larger than for the other models considered in this work.  
    }
	\label{fig:PXP_eigs}
\end{figure}

We can now use the value of $\mu=0.05$ to study the dynamics from special initial states in  Fig.~\ref{fig:PXP_init}. While for a given system-size the initial states display large variations in the late-time value of $\Delta^{(k)}$, for all of them we recover a clear decay of that value  with $N_B$. This suggests that, in the thermodynamic limit, all states eventually deep-thermalize, regardless of  scarring. Nonetheless, even for randomly sampled states, it is apparent that the time needed to deep-thermalize is remarkably long:  the time is on the order of $10^3$, despite the microscopic energy scale of the Hamiltonian being $O(1)$. We note, however, that this thermalization time scale is highly sensitive to the type of perturbation added to the PXP model to break the anticommutation with $\hat Z$. As shown in Appendix~\ref{sec:PXP_pert}, another perturbation with a similar strength leads to the a much shorter timescale for thermalization, in line with that in the East model.

The behavior of $\Delta^{(k)}$ for the eigenstates of the PXP model, shown in Fig.~\ref{fig:PXP_eigs}, also reveals some non-thermalizing features. We find a very broad distribution of values even at infinite temperature. This is similar to what is seen in the entanglement entropy of eigenstates in this model~\cite{Khemani2018,Turner2018b}, due to the presence of many non-thermal eigenstates beyond the ``obvious'' QMBS outliers with low entanglement entropy. In this sense, the signs of ergodicity breaking  in higher moments ($k{>}1$) do not appear to provide additional information compared to 
the $k{=}1$ case, accessible in ETH.

\begin{figure}[t!]
	\centering
	\includegraphics[width=\linewidth]{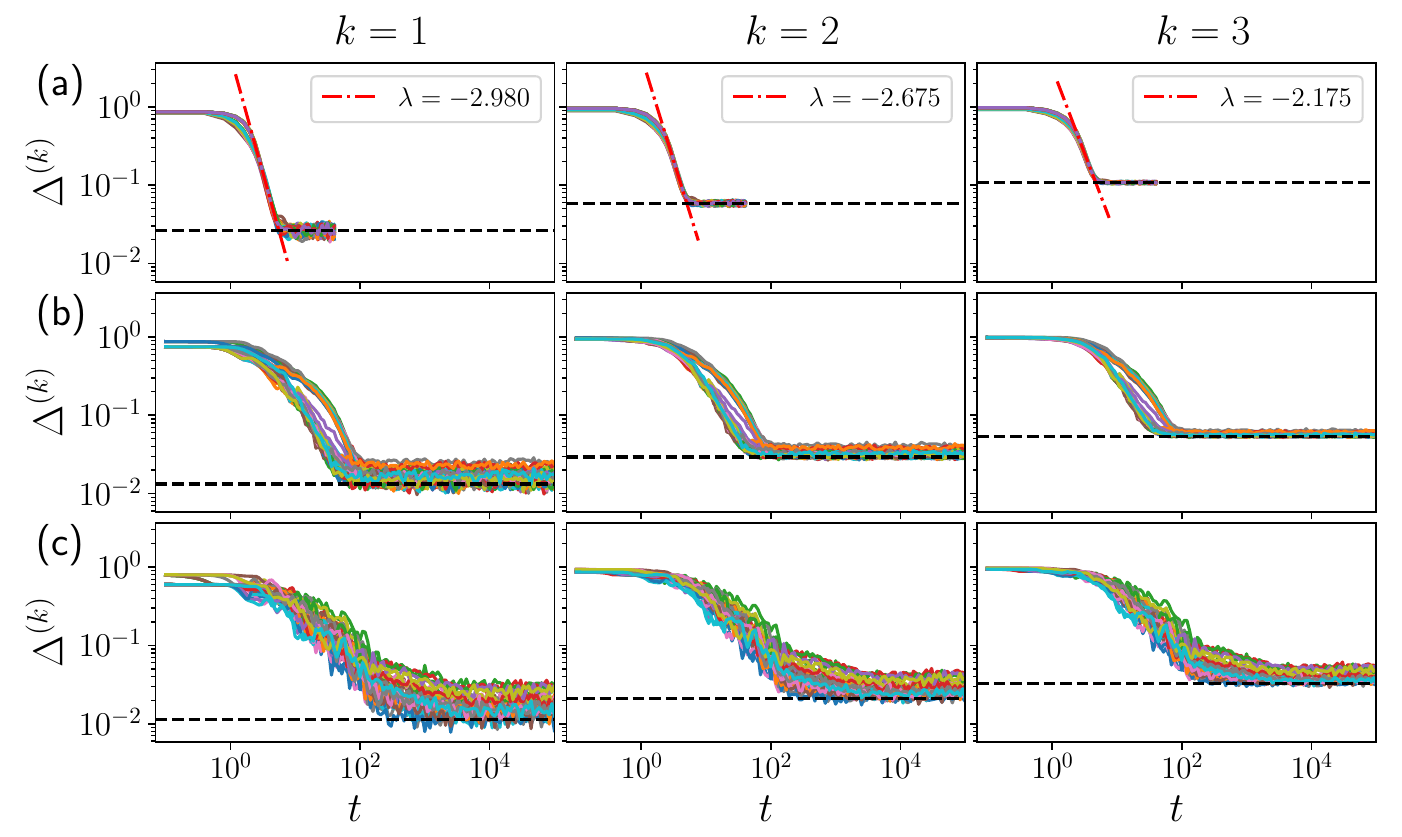}
	\caption{A comparison of $\Delta^{(k)}$ for (a) the SYK model with $N=16$, (b) the East model with $N=16$, and (c) the PXP model with $N=22$. Panel (a) also shows a power-law fit for the SYK model of the form $\Delta^{(k)}\sim t^{\lambda} $. For the latter two models, the initial state is a superposition of two states with $\alpha{=}\pm 1$, respectively. The difference in timescale as well as the variability between initial states is clearly visible between the three models.}
	\label{fig:comp_all}
\end{figure}

\section{Conclusions and discussion}\label{sec:conclusion}

In this work, we have studied deep thermalization for several models ranging from the maximally chaotic SYK to the heavily constrained PXP. 
Our results shed light on how, beyond usual symmetries, the invariance under time reversal and the existence of operators that anticommute with the Hamiltonian can hinder deep thermalization in otherwise chaotic models. As these properties do not influence the first $k{=}1$ moment of the projected ensemble, they have no effect on conventional ETH and thermalization of expectation values of local observables. This highlights the sensitivity of deep thermalization to special properties of the model beyond global symmetries. However, 
once the symmetries and anticommuting operators are properly accounted for, we find an exponential decay of $\Delta^{(k)}$ at late times for \emph{all} studied models and initial states in the numerically-accessible system sizes. Nonetheless, the rate of convergence to the Haar ensemble can vary greatly between different models or even between initial states in the same model, see Fig.~\ref{fig:comp_all} for a brief summary and comparison.  We find that for the SYK model, the decay of $\Delta^{(k)}$ with time is well fitted by a power-law for all $k$, as found in the Ising model in Ref.~\cite{Cotler2023}. However, we find a decay as $t^{-2.9}$ for $k=1$  against $t^{-1.2}$ in the Ising case, showcasing the faster thermalization in SYK. For the East and PXP models, while the decay also resembles a power-law, the variability between initial states implies that using a single exponent for all initial states is not meaningful. Overall, we find values similar to those of the Ising case for the East model, while for PXP the exponent is approximately halved.

The dependence of the exponent on the initial state is the most salient in the PXP model, where we also find that the time needed to reach the plateau is at least an order of magnitude larger than in other models considered. We note that this convergence rate has also been recently shown to sensitively depend on the boundary conditions~\cite{shrotriya2023nonlocality}.  While the convergence to the Haar ensemble in the PXP model can be enhanced via weak perturbations, its slowness reveals the presence of anomalous dynamics which affects a large part of the Hilbert space (and not just a few specific initial states) and persists up to surprisingly long times. It would be interesting to see if these results could explain the anomalous energy  transport in the PXP model at infinite temperature~\cite{Ljubotina2023}, where superdiffusion has been observed on time scales $\sim 10^2$ accessible to large-scale tensor network simulations. The onset of deep thermalization at even later times $\sim 10^3$ found here suggests that the observed superdiffusion may be transient before it gives way to diffusion. Nevertheless, this would still leave open the question of what physically sets such a long timescale for deep thermalization. The understanding of deep thermalization for initial conditions corresponding to a finite-temperature density matrix and the possible generalization of the projected ensemble to such cases~\cite{Jozsa1994} may shed light on this question. Furthermore, a more systematic investigation of models with variable constraints, proposed in Refs.~\cite{DooleyHypergrid,Desaules2022}, might yield more quantitative insights on the interplay between deep thermalization and dynamical connectivity of the Hilbert space, influenced by kinetic constraints. 

\begin{acknowledgments}
We would like to thank Shane Dooley and Silvia Pappalardi for useful discussions.  
This work was supported by the Leverhulme Trust Research Leadership Award RL-2019-015. J.-Y.D. acknowledges support by EPSRC grant EP/R513258/1. Statement of compliance with EPSRC policy framework on research data: This publication is theoretical work that does not require supporting research data.
This work was made possible by Institut Pascal at Universit\'e Paris-Saclay with the support of the program ``Investissements d’avenir'' ANR-11-IDEX-0003-01.
\end{acknowledgments}

\appendix

\section{Hyperplanes in Hilbert space}\label{sec:hyperp}

\begin{figure*}[ht!]
	\centering
	\includegraphics[width=0.8\linewidth]{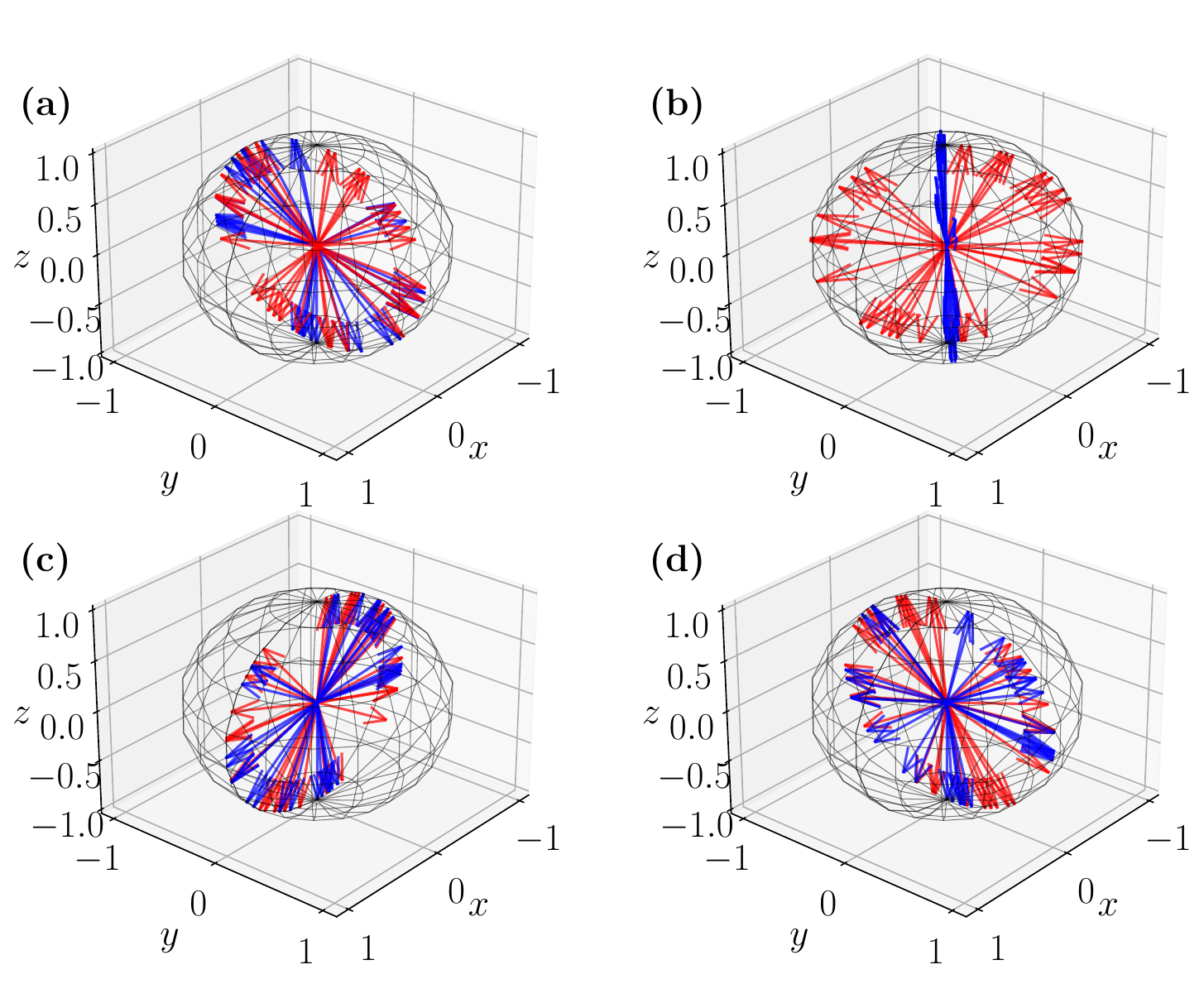}
	\caption{Bloch sphere representations of states in the projected ensemble for the rotated Ising model with $N=12$ and $N_A=1$. The rotation angles are (a) 0, (b) $-\pi/8$, (c) $-\pi/4$ and (d) $-\pi/2$. The states in red are linked to $z_B$ with $\alpha_{z_B}=+1$, while the blue ones are linked to $\alpha_{z_B}=-1$. The opposite rotation direction between the two is clear in panel (b).
    }
	\label{fig:Ising_hyper}
\end{figure*}

In this appendix, we illustrate the effects of change of basis matrices $\hat{V}$ that cannot simply be written as $\hat{V}=\hat{V}_A\otimes \hat{V}_B$ but instead as $\hat{V}=\sum_j \hat{V}^j_A\otimes \hat{V}^j_B$. For this purpose, we focus on an eigenstate at the middle of the spectrum of the Ising model as defined in Eq.~(\ref{eq:Ising}), with change of basis generated by 
\begin{equation}
    \hat{V}_{\phi}=\exp\left(i\phi\prod_j \sz_j \right),
\end{equation}
with various values of $\phi$. Let us denote by $\ket{E_0}$ the real eigenstate we study when $\hat{V}=\mathbb{1}$. Then, for any $\hat{V}_{\phi}$, the same eigenstate becomes $\ket{E_\phi}=\hat{V}_{\phi}\ket{E}$, as $\hat{V}_{\phi}$ is unitary. We note that it can be rewritten as $\hat{V}_{\phi}= \cos(\phi)\mathbb{1}+i\sin(\phi)\prod_j \sz_j$. This allows us to rewrite $\ket{E_\phi}$ using the same steps as in  Eq.~(\ref{eq:rot_ZA}). We also set $N_A=1$ to finally get
\begin{equation}
        \ket{E_{\phi}}=\sum_{z_B}\sqrt{p_{z_B}}\exp\left[i\phi \alpha_{z_B}\sz_1\right]\ket{\psi_{z_B}^A}\otimes \ket{z_B}.
\end{equation}
As discussed in the main text, the rotation applied to the states in the projected ensembles are now dependent on the eigenvalue $\alpha_{z_B}=\pm 1$ of the $\ket{z_B}$ under $\prod_{j=2}^N \sz_j$. In Fig.~\ref{fig:Ising_hyper}, we plot a few important cases to consider.

If $\phi=0$, there is no rotation and the all $\ket{\psi^A_{z_B}}$ lie in the $XZ$ plane as they are real (up to an overall phase). If $\phi$ is real and not a multiple of $\pi/4$, then the $\ket{\psi^A_{z_B}}$ with $\alpha_{z_B}=\pm 1$ lie in different planes as they get rotated in opposite directions. If $\phi=\pm \pi/2$, states for both values of $\alpha_{z_B}=\pm 1$ get rotated by a half circle. Because of this the sign of $\alpha_{z_B}$ is irrelevant and the dependence on $z_B$ is lost. This means that the projected ensemble is affected by an overall rotation, leading the \emph{exact} same $\Delta^k$ as for $\phi=0$. This is also true for all multiple of $\pi/2$. Finally, we consider the case $\phi=\pm \pi/4$. This is the most important case as it corresponds to the angle for time-evolved states in the presence of time-reversal and an ``antisymmetry''. For this angle, both values of $\alpha_{z_B}$ also end up on the same plane (the $YZ$ plane here). However, in this case the value of  $\alpha_{z_B}$ still matters and thus this transformation is not equivalent to a global rotation. Indeed, let us consider a vector pointing in the $+Y$ direction. In the case of $\phi=-\pi/4$, for $\alpha_{z_B}=+1$ it will end up pointing in the $-Y$ direction, while for  $\alpha_{z_B}=-1$ it would be in the $+Y$ direction. As a consequence, the $\Delta^{(k)}$ value will \emph{not} be identical after this rotation. However, as the projected ensemble is still contained in a single hyperplane, $\Delta^{(k)}$ should fluctuate around the same lower limit as for $\phi=0$ for large enough $N_B$.

\section{Perturbations of the PXP model}\label{sec:PXP_pert}

\begin{figure*}[ht!]
	\centering
	\includegraphics[width=0.8\linewidth]{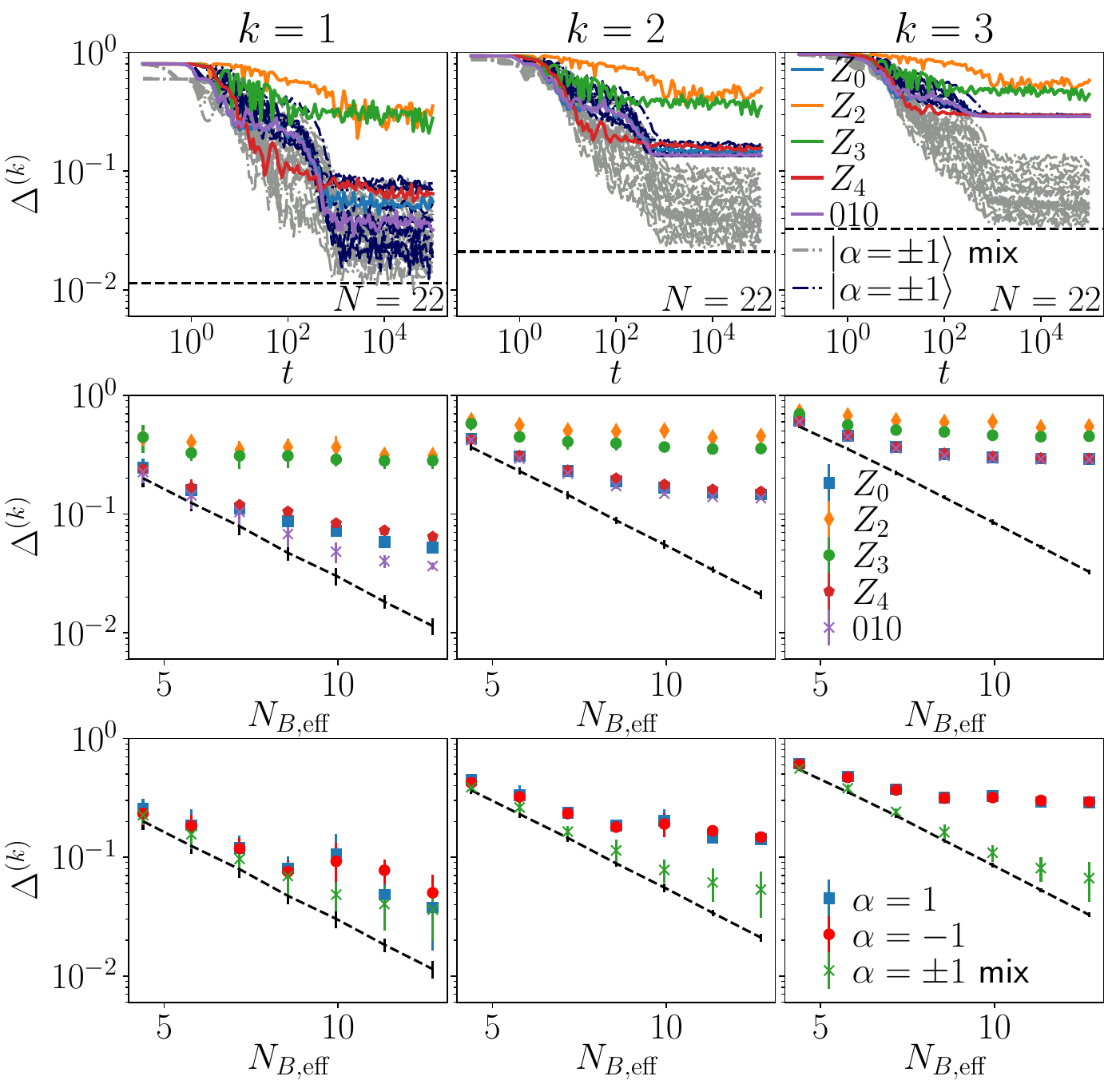}
	\caption{$\Delta^{(k)}$ for time-evolved states in the PXP model with the PXPZ perturbation in Eq.~(\ref{eq:PPXP}) and perturbation strength $h=0.024$. As the perturbation does not break the ``antisymmetry'', we still see a strong difference between states with $\alpha=\pm 1$ and their superposition. However, even for the latter, we find strong variations between initial states, already visible at $k=1$.
    }
	\label{fig:PXPZ}
\end{figure*}

Instead of chemical potential, used in the main text, thermalization properties of the PXP model can be tuned by applying other types of perturbations, two of which are considered in this Appendix. The first perturbation we use is 
\begin{equation}\label{eq:PPXP}
    \hat{H}\!=\!\hat{H}_\mathrm{PXP}-h\hat{H}_\mathrm{int}, \ \hat{H}_\mathrm{int}\!=\!\!\sum_j \Pz_{j-1} \sx_j\Pz_{j+1}\left(\sz_{j-2}+\sz_{j+2}\right).
\end{equation}
This perturbation has been shown to enhance the scarring dynamics from the $\ket{\mathbb{Z}_2}$ state at $h{\approx} 0.05$~\cite{Choi2018}. For $h{\approx} 0.024$, it instead pushes the level statistics of the model closer to the Poisson ensemble~\cite{Khemani2018}. While both effects are most prominent for PBCs, we still see the same qualitative behavior with OBCs. Here we will focus on the latter value $h{\approx} 0.024$, as it affects the entire spectrum. The results of $\Delta^{(k)}$ for time-evolved states with this perturbation are shown in Fig.~\ref{fig:PXPZ}
The special states $\ket{\mathbb{Z}_2}$ and $\ket{\mathbb{Z}_3}$ show a larger value for their $\Delta^{(k)}$. Meanwhile, as expected, we also see worse thermalization for all states, with a very high variability between them. This effect is prominent for all $k$. 

\begin{figure*}[ht!]
	\centering
	\includegraphics[width=0.8\linewidth]{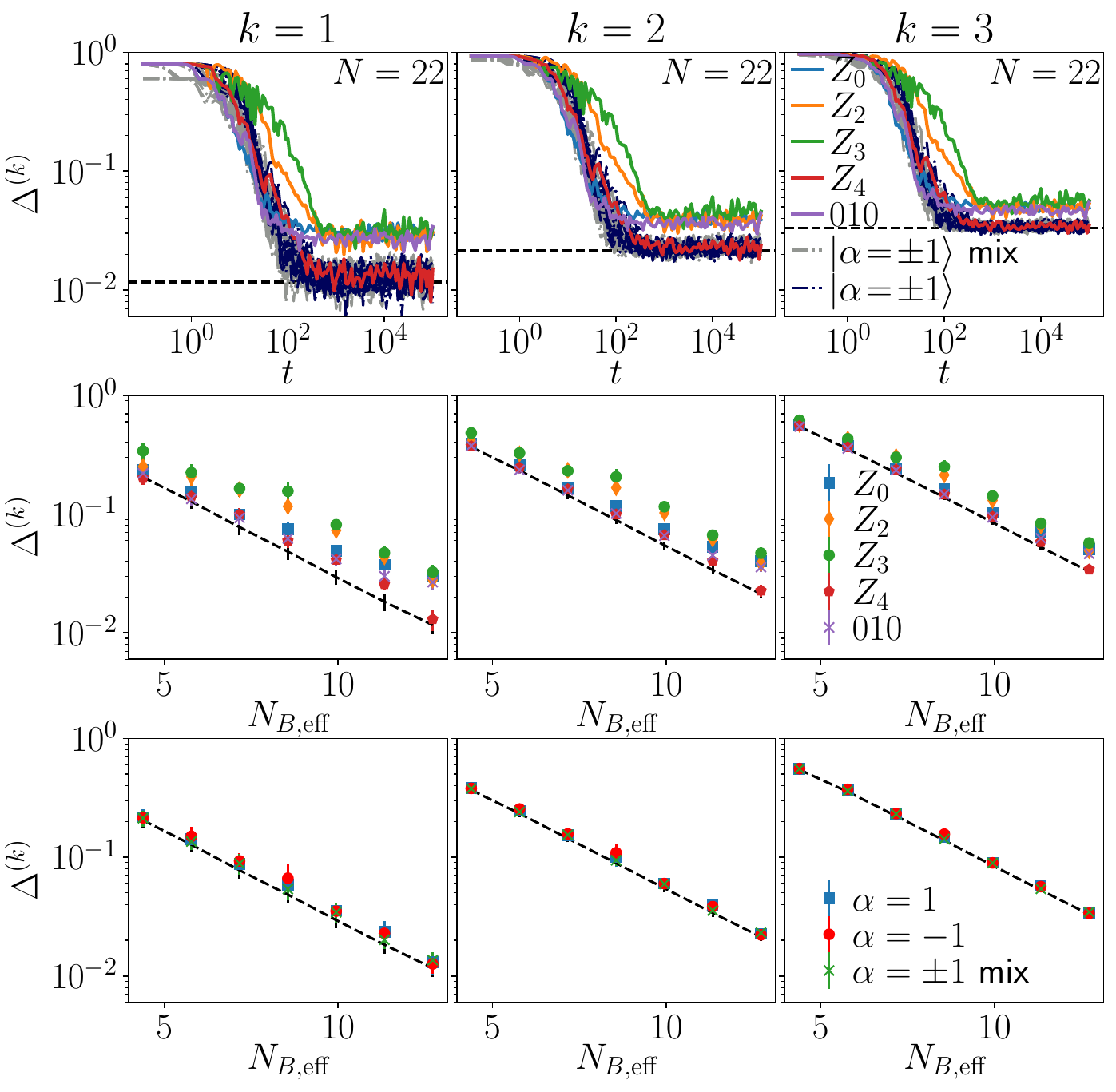}
	\caption{$\Delta^{(k)}$ for time-evolved states in the PXP model with the PXPXP perturbation in Eq.~(\ref{eq:PXPXP}) and perturbation strength $\lambda=0.05$. Even such a weak perturbation is sufficient to significantly lower the $\Delta^{(k)}$ plateau for all state while also shortening the time it takes to reach it.
    }
	\label{fig:PXPXP}
\end{figure*}

The second perturbation has an opposite effect, making the entire spectrum more thermalizing. The perturbation was first discussed in Ref.~\cite{Turner2018b} and is defined as 
\begin{equation}\label{eq:PXPXP}
\hat{H}\!=\!\hat{H}_\mathrm{PXP}+\lambda\hat{H}_\mathrm{therm}, \ \hat{H}_\mathrm{therm}\!=\!\sum_j \Pz_{j-2} \sx_{j-1}\Pz_{j}\sx_{j+1}\Pz_{j+2}.
\end{equation}
We show its effect in Fig.~\ref{fig:PXPXP} for $\lambda=0.05$. Even such a small perturbation strength is enough to suppress variations of $\Delta^{(k)}$ between initial states. However, we still see a somewhat higher $\Delta^{(k)}$ for the $\ket{\mathbb{Z}_2}$ and $\ket{\mathbb{Z}_3}$ states and thus a stronger perturbation is still needed to completely destroy scarring. Another, very clear, effect of this perturbation is the reduction of the timescale needed to reach the plateau of $\Delta^{(k)}$, which goes from $t\approx 10^3$ down to $t\approx 10 ^2$ upon the addition of this perturbation. The latter timescale is similar to that observed in the East model in Sec.~\ref{sec:PX}. 

\bibliography{biblio.bib} 

\end{document}